\documentclass[aps,prd,groupaddress,floatfix,tighten,nofootinbib,showpacs,
amsfonts,superscriptaddress]{revtex4}

\usepackage{url}
\usepackage{xcolor}
\usepackage{hyperref}
\colorlet{shadecolor}{gray!15}
\definecolor{greenLinks}{rgb}{0,0.6,0}
\definecolor{blueLinks}{rgb}{0,0,0.6}
\definecolor{redLinks}{rgb}{0.6,0,0}
\definecolor{tempText}{rgb}{0.55,0.10,0.67}
\definecolor{eprintLinks}{rgb}{0.4,0.4,0.4}
\definecolor{journalLinks}{rgb}{0.6,0,0}
\DeclareMathSymbol{\not}{\mathrel}{symbols}{"36}
\usepackage{amssymb}
\usepackage{amsmath,color}
\usepackage{epsfig}
\usepackage{graphicx,epsf,epsfig}
\usepackage{footnote}
\usepackage{multirow}
\usepackage{graphicx}
\usepackage{subfigure}
\usepackage{lipsum}
\usepackage{multirow} 
\usepackage{tabulary}
\usepackage{rotating}
\usepackage{array}
\usepackage[normalem]{ulem}
\usepackage{color}
\def\slc#1{
 \setbox0=\hbox{$#1$}                      
 \dimen0=\wd0                              
 \setbox1=\hbox{/} \dimen1=\wd1            
 \ifdim\dimen0>\dimen1                     
 \rlap{\hbox to \dimen0{\hfil/\hfil}}      
 #1                                        
 \else                                     
 \rlap{\hbox to \dimen1{\hfil$#1$\hfil}}   
 /                                         
\fi}

\def\be{\begin{equation}}
\def\ee{\end{equation}}
\def\gs{\mathrel{\rlap{\raise 0.511ex \hbox{$>$}}{\lower 0.511ex \hbox{$\sim$}}}}
\def\ls{\mathrel{\rlap{\raise 0.511ex \hbox{$<$}}{\lower 0.511ex \hbox{$\sim$}}}}

\newcommand{\ba}{\begin{array}{c}}
\newcommand{\baz}{\begin{array}{cc}}
\newcommand{\barrr}{\begin{array}{rrr}}
\newcommand{\bad}{\begin{array}{ccc}}
\newcommand{\bav}{\begin{array}{cccc}}
\newcommand{\baf}{\begin{array}{ccccc}}
\newcommand{\bea}{\begin{equation} \begin{array}{c}}
\newcommand{\eea}{ \end{array} \end{equation}}
\newcommand{\ea}{\end{array}}

\usepackage[normalem]{ulem}

\def\21{$\mathrm{SU(2)_L \otimes U(1)_Y}$ }

\newcommand{\ignore}[1]{}

\usepackage{float}
\usepackage{cleveref}
\newcommand{\bes}{\begin{subequations}}
\newcommand{\ees}{\end{subequations}}

\begin{document}

\title{Linking LFV Higgs decays $h\to \ell_i \ell_j$ with CP violation in multi-scalar models}

\vspace*{-15mm}
\begin{flushright}
CIFFU 17-01
\end{flushright}
\vspace*{0.7cm}

%
\allowdisplaybreaks \allowdisplaybreaks[2]
\newcommand{\AddrCIFFU}{Centro de Internacional de F\'{\i}sica Fundamental, 
  Benem\'erita Universidad Aut\'onoma de Puebla.}
 \newcommand{\AddrFCFMBUAP}{Fac. de Cs. F\'{\i}sico Matem\'aticas, 
  Benem\'erita Universidad Aut\'onoma de Puebla,\\
  Apdo. Postal 1152, Puebla, Pue.  72000, M\'exico.}
 \newcommand{\AddrFCEBUAP}{Fac. de Cs. de la Electr\'onica, 
  Benem\'erita Universidad Aut\'onoma de Puebla,\\
  Apdo. Postal 542, Puebla, Pue. 72000, M\'exico.}
 
%

\author{E. Barradas-Guevara}
 \email{barradas@fcfm.buap.mx}
  \affiliation{\AddrCIFFU}
 \affiliation{\AddrFCFMBUAP}

%
%
%
\author{J. Lorenzo Diaz-Cruz}
 \email{jldiaz@fcfm.buap.mx}
  \affiliation{\AddrCIFFU}
 \affiliation{\AddrFCFMBUAP}
 %
 %
\author{O. F\'elix-Beltr\'an}
 \email{olga.felix@correo.buap.mx}
 \affiliation{\AddrCIFFU}
 \affiliation{\AddrFCEBUAP}
%
%
\author{U. J. Saldana-Salazar}
 \email{usaldana@fcfm.buap.mx}
  \affiliation{\AddrCIFFU}
 \affiliation{\AddrFCFMBUAP}
%
\date{\today}

\begin{abstract} \vspace{.3 cm}
The study of LFV decays of the Higgs boson,  $h\to \ell_i \ell_j$, has become an active research subject both from the 
experimental and theoretical points of view. 
Such decays vanish within the SM and are highly suppressed in several theoretical extensions.  
Due to its relevance and relative simplicity to reconstruct the signal at future colliders, it is an important tool to probe 
SM extensions where it could reach detectable levels. Here we identify a mechanism that induces LFV Higgs interactions, 
by linking it with the appearance of CP violation in the scalar sector, within the context of general multi-Higgs models. 
We then focus on the simplest model of this type to study its phenomenology. The scalar sector of this minimal model 
consisting of a Higgs doublet and a Froggatt--Nielsen (FN) (complex) singlet is studied thoroughly, including 
the scalar spectrum and the Yukawa interactions. Constraints on the parameters of the model are derived from low-energy
observables and LHC Higgs data, which is then applied to study the resulting predicted rates for the decay 
$h\rightarrow \tau \mu$. 
Overall, branching ratios for $h \rightarrow \tau \mu$ of the order $10^{-3}$ are obtained within this approach
consistent with all known constraints.
\end{abstract}
\maketitle
\newpage
\baselineskip 24pt

\clearpage
\setcounter{page}{1}

\section{Introduction\label{Sec:Intro}}
Finding some signal of New Physics (NP) {has} been majorly expected for long time, 
specially, after the discovery at the 
LHC of a Higgs-like particle with mass, $m_h=125.09\; \pm0.21 \;(\text{stat.}) \; \pm 0.11\;(\text{syst.})$ 
GeV~\cite{higgs-atlas:2012gk,higgs-cms:2012gu,Aad:2015zhl}. 
But an scenario with no new findings at the  the LHC 
portraits an apocalyptic future, which gets reinforced
after current measurements of the spin, parity, and couplings,
of the {  newly found} boson, seem  consistent with 
the Standard Model (SM) prediction~\cite{Gunion:1989we}.  However, the existence of a light Higgs boson seems
problematic (\textit{i.e.} the hierarchy problem) and calls for NP. Similarly, the SM has other
{theoretical} open issues, such as the flavor {problem}, unification, etc.~\cite{Pomarol:2012sb,Martin:1997ns}, 
which also motivate NP models. 

Many papers have been devoted to study the pattern of Higgs couplings from the LHC data, for instance~\cite{Espinosa:2012ir,Giardino:2013bma}. 
The couplings of the Higgs particle to a pair of massive gauge bosons or fermions 
are proportional  to the particle mass.  However,  the LHC has tested only a few of these couplings,
\textit{i.e.} the ones with the heaviest SM fermions and the $W$ and $Z$ bosons, while non-standard Higgs couplings, 
including the flavor violating (FV) ones, are predicted in many models of physics beyond the SM and they  
have been tested  at the LHC only recently~\cite{Branco:2011iw, DiazCruz:2004tr, DiazCruz:2002er}.

The scalar sector is certainly playing a big role within the flavor problem. 
Already a dimension six operator can easily introduce Flavor Changing Neutral Currents (FCNC's),
\begin{eqnarray}
	-{\cal L}_Y^{\text{NP}} \subset \frac{\lambda_{ij}}{\Lambda^2} \bar{F}_i f_j \Phi (\Phi^\dagger \Phi) ,
\end{eqnarray}
as $\lambda_{ij}$ is not simultaneous diagonalized by the same unitary transformations which
bring the Yukawa matrices to the mass basis (diagonal form). That is, the initial Yukawa matrices can effectively get new contributions with rather major consequences. And, on the other hand, if we consider multi-Higgs models, the simplest case, that is a second Higgs doublet, could also immediately bring about the same scenario. Therefore, a huge part of the flavor problem could be arising from the still unknown scalar sector.

In other words, among the sectors of the SM, the one that is equally or even less  understood is the Yukawa sector. As most of the SM arbitrariness (parameters) is precisely emerging from it. In fact, the flavor problem originates from all the phenomenological observed patterns in fermion masses and mixings, which get produced from the Yukawa couplings. In this sense, a thorough understanding of the Yukawa couplings would then mean a big step to the solution of the flavor problem, {see for example the following idea~\cite{Saldana-Salazar:2015raa,Saldana-Salazar:2016hxb,DiazCruz:2004ss}}.

From a phenomenological point of view, the smallness of neutrino masses allows the consideration of an approximate conservation of lepton flavor numbers. In order to see this, recall that
the kinetic part of the SM Lagrangian has an accidental and global flavor symmetry group given by,
\begin{eqnarray}
	{\cal G}_{SM} = U(3)^Q_L \times U(3)^u_R \times U(3)^d_R \times U(3)^E_L \times U(3)^e_R, 
\end{eqnarray}
which, after consideration of the Yukawa Lagrangian, gets broken to,
\begin{eqnarray}
	{\cal G}_{SM} \rightarrow U(1)_B \times U(1)_{L_e} \times U(1)_{L_\mu} \times U(1)_{L_\tau}.
\end{eqnarray}
This remnant is then associated to the conservation of baryon number ($B$) and three different lepton numbers ($L_\alpha$, $\alpha = e,\; \mu,\; \tau$). It is in this sense that lepton flavor is thus defined. Now, through a different choice
of basis we may write the latter in the following manner~\cite{Heeck:2016xwg},
\begin{eqnarray}
	U(1)_{B-L} \times U(1)_{B+L} \times U(1)_{L_\mu - L_\tau} \times U(1)_{L_\mu + L_\tau - 2L_e},
\end{eqnarray} 
where we have defined total lepton number as $L=L_e + L_\mu + L_\tau$. 
For several reasons, it is useful to express it in this way as on one hand the first factor,
$U(1)_{B-L}$, is the only conserved part by non-perturbative processes at the quantum level while on the
other, a model independent approach can easily lead by this construction to identify which
lepton-flavor-violating processes are required to establish that the entire flavor group is broken~\cite{Heeck:2016xwg}. We already know that the addition of neutrino masses breaks
lepton number ($L_\alpha$), however, their smallness allows the consideration that the left symmetry group is an
excellent approximate symmetry for charged leptons. Therefore, the observation of charged lepton-flavor-violating transitions would imply physics beyond the SM~\cite{Heeck:2016xwg}.

 Several ideas have been proposed to {address} the flavor problem~\cite{Isidori:2010kg}, for instance: 
textures~\cite{Fritzsch:1977vd,Fritzsch:1977za,Fritzsch:1979zq},  GUT-inspired relations~\cite{Georgi:1979df,King:2013hj}, flavor symmetries~\cite{Fritzsch:1999ee,Ishimori:2010au}, {hierarchical mass ratios~\cite{Hollik:2014jda,Saldana-Salazar:2016pms}}, {multi-Higgs doublet models~\cite{Blechman:2010cs}}, and radiative generation of fermion masses~\cite{Balakrishna:1988ks,Babu:1989fg,Ma:1998dn,DiazCruz:2005qz}. 
The flavor symmetry approach can be supplemented with the {Froggatt}-Nielsen (FN) mechanism~\cite{Froggatt:1978nt}, which
assumes that above some scale $M_F$ such symmetry forbids the appearance of the Yukawa couplings; here, the
SM fermions are charged under this symmetry (which could be of the Abelian type $U(1)_F$).
Nonetheless, the Yukawa matrices can arise  through non-renormalizable operators. 
The Higgs spectrum of these models could include light flavons, which could then mix with
the Higgs boson, {see for example~\cite{Diaz-Cruz:2014pla}}. 

In these models, the diagonal Flavor Conserving (FC) couplings of the light SM-like Higgs boson 
could deviate from the SM, while FV couplings could be induced at small rates too~\cite{Diaz-Cruz:2014pla}, 
but still at rates that could produce detectable signals. 
 On the other hand, extending the Higgs sector of the SM, offers the possibility to include
an Scalar Dark Matter candidate, as it is the case of the well studied Inert Doublet Model~\cite{Bonilla:2014xba}. 
There are relevant motivations to supplement this model with a complex singlet, for instance to have extra sources
of CP violation, as in the {Inert Dark Matter model with a complex Singlet} (IDMS) studied recently~\cite{Bonilla:2014xba}.

An interesting probe of FV Higgs couplings is provided by the decay $h \to \tau \mu$, 
which was initially studied in refs.~\cite{Pilaftsis:1992st, DiazCruz:1999xe}. 
Such decay vanish within he SM and is suppressed in some extensions, however, their relevance motivates 
looking for extensions where it could be detectable. Subsequent studies on 
detectability of the signal appeared soon after~\cite{Han:2000jz, Assamagan:2002kf, Kanemura:2005hr}. 
Precise loop calculations with massive neutrinos, SUSY, and other models appeared in~\cite{Arganda:2004bz,DiazCruz:2002er,Brignole:2004ah,DiazCruz:2008ry}.
The recent search for this decay at the LHC~\cite{Khachatryan:2015kon}, {has} resulted in a bound
for the corresponding branching ratio of order $Br(h\to \tau\mu) < 1.51 \times 10^{-2}$  at 95\% C.L..
Furthermore, given that the best fit to the data gave $Br(h\to \tau\mu) = 0.84^{+0.39}_{-0.37} \times 10^{-2}$ (recent results from the LHC has reduced this value to $0.55^{+0.33}_{-0.32} \times 10^{-2}$ \cite{Aad:2016blu}),
many more papers appeared afterwards, trying to explain this result~ \cite{Das:2015zwa,Varzielas:2015joa,Baek:2015fma,
Baek:2015mea,Vicente:2015cka,Baek:2016pef,Baek:2016kud}. 
Nevertheless, the search for this lepton flavor violation (LFV) Higgs decay could be one great opportunity to find new physics at the LHC Run II.

In this paper,  we study LFV Higgs decays and identify a mechanism that induces LFV Higgs 
interactions for the light SM-like Higgs boson, by linking it with the phenomena of CP violation within the context of multi-Higgs doublet models with an extra FN singlet. 
We then provide a simple model to study its phenomenology, namely, a model with an scalar sector consisting of a Higgs doublet 
and a FN singlet, where the neutral component of the doublet mixes with the imaginary component of the FN singlet.  

The organization of this paper goes as follows. After some introductory ideas (Section \ref{Sec:Intro}), we present 
in Section \ref{Sec:NHDM} the realization of
our mechanism within the context of a $N$-Higgs doublet model with one FN singlet. Afterwards, in Section \ref{Sec:Model}, 
we consider its simplified version, namely the one with one Higgs {doublet} and one FN singlet. Then, 
in Section \ref{Sec:Matrices}, we study the Higgs potential of the model and find out that one of the extra mass eigenstates 
tends to be lighter, with a mass of order of the light SM-like Higgs mass.
Then, it is shown how the mixing of the real part 
of the doublet with the imaginary component of the FN singlet induces sizeable  LFV Higgs couplings of the light physical
Higgs boson, which can have large LFV decays. 
Also, the couplings to the SM-like Higgs boson are studied in Section \ref{Sec:Higgs-constraints}, including
the low-energy constraints.  Within this section, 
the Higgs decays are computed and discussed, while
the evaluation of branching ratios for the LFV Higgs decays is also presented, as well as its comparison with bounds 
from the LHC collaborations.  Conclusions are given in Section \ref{Sec:Conclusions}.

\section{LFV and CPV in a $N$-Higgs doublet model with one FN singlet\label{Sec:NHDM}}
We shall discuss the proposed scenario, within the context of a
model made out of $N$-Higgs doublets plus one complex singlet charged under a FN symmetry. We want to show that this study is, in fact, quite straightforward. Let us see this.

On one hand, we know that the analysis of the vacuum structure, which started with the work of~\cite{DiazCruz:1992uw}, may always be brought, by a smart choice of basis, to the equivalent case of either three or two Higgs doublets \cite{Barroso:2006pa}. This reduction means that if at tree level one has a normal minimum, in the former case, it cannot always be below any charge breaking stationary point, thus allowing violation of electric charge; whereas in the case of reaching an equivalent two Higgs doublet scenario, the normal minimum can always be found to lay below any other stationary point therefore leaving $U(1)_{\text{EM}}$ invariant \cite{Barroso:2005sm,Ferreira:2004yd}. Furthermore, the reduction of $N$ scalars to the study of three can still be cured and violation of electric charge can still be avoided by satisfying the following sufficient condition. Basically, this condition requests that the parameters of the potential are such that after arriving to the so called B-basis the normal vacuum structure mimics that of a two Higgs doublet model \cite{Barroso:2005sm}. 

On the other hand, the steps leading from the initial effective Yukawa Lagrangian to its form in the mass basis for both fermion and scalar fields are straightforward. In the following, we show this calculation. 

Let us consider an $N$-Higgs doublet model plus a FN field. The effective Yukawa Lagrangian takes the form,
	\begin{eqnarray}
		-{\cal L}_Y = \sum_{i,j,f,a} \alpha^{f,a}_{ij} \left( \frac{S_F}{\Lambda_F} \right)^{\kappa^{f,a}_{ij}}
		\bar{F}_{i} f_j {\Phi_a} + {\textrm{H.c.}},
	\end{eqnarray} 
where $i,j =1,2,3$, $f=u,d,e$, and $a=1,2,\dots,N$.	
Two immediate possibilities arise: not a single Higgs doublet is being shared by more than one fermion type with a 
given electric charge ($u$,$d$, $e$) or at least one Higgs doublet is being shared by more than one fermion 
type with a given electric charge. If a FN field was not included
then the former scenario would mean natural flavor conservation while the latter leads to flavor violation. 
Nevertheless, as we have included a singlet flavon scalar field flavor violation will be induced irrespective 
of the number of Higgs fields being shared among the different fermion types. Therefore, we will not focus 
here on this theoretical aspect which is then basically translated into what type of scalar doublet 
model we are building (which Higgs fields couple to which fermion fields) as in any case flavor is violated. 

The next step is a generic one. We write the flavon as, 
\begin{eqnarray}
 S_F = \frac{1}{\sqrt{2}} ({w} e^{i\xi}  + s_1 + i p_1 ),
\end{eqnarray}
and make the linear expansion,
\begin{eqnarray}
\left( \frac{S_F}{\Lambda_F} \right)^{\kappa^{f,a}_{ij}} \simeq 
\left( \frac{u}{\sqrt{2}\Lambda_F} \right)^{\kappa_{ij}^{f,a}} \left[ 1 +\kappa_{ij}^{f,a} \left( \frac{s_1 + i p_1}{u} \right)  \right],
\end{eqnarray} 
where we have denoted $u= w e^{i \xi}$. Then we expand the Yukawa Lagrangian,
	\begin{eqnarray}
		-{\cal L}_Y^f \simeq \sum_{i,j,f,a} Y^{f,a}_{ij} 
		\bar{F}_{i} f_j \Phi_a + \sum_{i,j,f,a} \kappa_{ij}^{f,a} Y_{ij}^{f,a} \frac{s_1}{u} \bar{F}_{i} f_j \Phi_a
		+ \sum_{i,j,f,a} i \kappa_{ij}^{f,a} Y_{ij}^{f,a} \frac{p_1}{u} \bar{F}_{i} f_j \Phi_a + {\textrm{H.c.}},
	\end{eqnarray}
where we have identified
	$Y_{ij}^{f,a} = \alpha_{ij}^{f,a} (u/ \sqrt{2}\Lambda_F)^{\kappa_{ij}^{f,a}}$. 
	
	The neutral component of the 
	Higgs fields can be written in {terms} of their vev,
	\begin{eqnarray}
		[\Phi_a]_0 = \frac{v_a + \phi_0^a + i\chi^a}{\sqrt{2}}.
	\end{eqnarray}
	Now,	after {the} substitution of the vevs we obtain,
	\begin{eqnarray}
		-{\cal L}_Y^f = \sum_{i,j,f} \left[ {\bf M}^f_{ij} 
		\bar{F}_{i} f_j  +  \sum_a \frac{{\bf M}^{f,a}_{ij}}{v_a} \bar{F}_{i} f_j \phi_0^a 
		+ \sum_a {\bf Z}_{ij}^{f,a} e^{-i\xi }\frac{s_1+ip_1}{w} \bar{F}_{i} f_j \frac{v_a + \phi_0^a}{\sqrt{2}}
		+  {\textrm{H.c.}} \right],
	\end{eqnarray}	
	where ${\bf M}_{ij}^{f,a} = \frac{v_a}{\sqrt{2}} Y_{ij}^{f,a}$, ${\bf M}_{ij}^f = \sum_a {\bf M}_{ij}^{f,a} $, ${\bf Z}_{ij}^{f,a} = \kappa_{ij}^{f,a} {Y}_{ij}^{f,a}$,
	and we have assumed the unitary gauge $G_z \rightarrow 0$, 
	and keep only the imaginary component of the
	scalar singlet. 
	
	We may now bring the fermion fields to the mass basis, implying by it,
	\begin{eqnarray}
		-{\cal L}_Y^f = \sum_{i,j,f} \left[ m^f_i \delta_{ij}
		\bar{F}'_{i} f'_j  +  \sum_a \frac{\widetilde{\bf M}^{f,a}_{ij}}{v_a} \bar{F}'_{i} f'_j \phi_0^a 
		+ \sum_a \widetilde{\bf Z}_{ij}^{f,a} e^{-i\xi }\frac{s_1+ip_1}{w} \bar{F}'_{i} f'_j \frac{v_a + \phi_0^a}{\sqrt{2}}
		+  {\textrm{H.c.}} \right],
	\end{eqnarray}
	where the tilde matrices are in general not diagonal. From this picture it becomes very
	apparent the existence of two sources of flavor violation: the one coming from 
	$\widetilde{\bf M}^{f,a}$ and only related to the fact of having multiple Higgs doublets and
	$\widetilde{\bf Z}^{f,a}$ which effectively entails the emergent interactions coming from
	the FN field. 
	
Further reduction may be achieved by assuming that the matrix parametrizing 
the interactions with the FN field is Hermitian, $\widetilde{\bf Z}^{f,a\, \dagger} = \widetilde{\bf Z}^{f,a}$. Obtaining only for that term,
\begin{eqnarray} \nonumber
\widetilde{\bf Z}_{ij}^{f,a} e^{-i\xi } \left( \frac{s_1+i p_1}{w} \right) \bar{F}'_{i} f'_j \left( \frac{v_a + \phi^a_0}{\sqrt{2}} \right)+ {\textrm{H.c.}} 
	= \widetilde{\bf Z}_{ij}^{f,a} \left[ \bar{\cal F}'_{i} {\cal F}'_j (s_1 \cos \xi + p_1 \sin \xi )
+  {i} \bar{\cal F}'_{i} \gamma_5 {\cal F}'_j (- s_1 \sin \xi + p_1 \cos \xi ) \right] \frac{v_a + \phi^a_0}{w\sqrt{2}},	
\end{eqnarray}
where $P_{L} {\cal F} = F$, $P_R {\cal F} = f$, and $P_{L(R)} = \displaystyle{\frac{1\mp \gamma_5}{2}}$. Its substitution then means,
\begin{eqnarray} \nonumber
		-{\cal L}_Y^f = \sum_{i,j,f} \left[ m^f_i \delta_{ij}
		\bar{F}'_{i} f'_j  +  \sum_a\frac{\widetilde{\bf M}^{f,a}_{ij}}{v_a} \bar{F}'_{i} f'_j \phi_0^a + {\textrm{H.c.}}
		+ \sum_a \widetilde{\bf Z}_{ij}^{f,a} \left[ \bar{\cal F}'_{i} {\cal F}'_j (s_1 \cos \xi + p_1 \sin \xi ) \right. \right.
+ \\ \left. \left. {i} \bar{\cal F}'_{i} \gamma_5 {\cal F}'_j (- s_1 \sin \xi + p_1 \cos \xi ) \right] \frac{v_a + \phi^a_0}{w\sqrt{2}} \right],
\end{eqnarray}

Finally, we must bring the scalar fields to their mass basis by means of the following orthogonal transformation, 
\begin{eqnarray}
		\begin{pmatrix}
			\phi_0^1  \\
			\phi_0^2 \\
			\vdots \\
			\phi_0^N \\
			s_1 \\
			p_1 
		\end{pmatrix}
		= \begin{pmatrix}
			{\cal O}_{1k} h_k \\
			{\cal O}_{2k} h_k \\
			\vdots	\\
			{\cal O}_{(N+2)k} h_k
		\end{pmatrix},
\end{eqnarray}
where $h_k$ ($k=1, \dots, N+2$) are the mass eigenstates.
The scalar potential dictates the mixing pattern of the Higgs and flavon fields. When an accidental symmetry is broken 
spontaneously, there appears a pseudo-Goldstone boson, which would be the lightest flavon. The Higgs particles of the 
spectrum would then mix with the flavons. If CP is conserved, the real (imaginary) components of the Higgs and flavons
would mix. When CP is violated, it is possible to induce mixing among the Higgs and the imaginary components of the flavons. 
Given the possibility to study LFV couplings of the 125 GeV Higgs, we would prefer  to have optimal rates. 
Therefore, we shall admit the possibility that CP is violated, focusing then in the mixing of the light SM-like Higgs 
and the imaginary component of the flavon, which will be in general the lightest flavon state, which would then, 
offer the possibility of inducing larger rates for LFV Higgs decays.
	
In the Higgs mass basis,
\begin{eqnarray} \label{Eq16}
\begin{split}
		-{\cal L}_Y^f = \sum_{i,j,f} \left[ m^f_i \delta_{ij}
		\bar{F}'_{i} f'_j  + \sum_{a,k} \frac{\widetilde{\bf M}^{f,a}_{ij}}{v_a} \bar{F}'_{i} f'_j [{\cal O}_{ak} h_k] + {\textrm{H.c.}} 
		+ \sum_{a,k} \widetilde{\bf Z}_{ij}^{f,a} \left[ \bar{\cal F}'_{i} {\cal F}'_j ([{\cal O}_{(N+1)k} h_k]  \cos \xi + [{\cal O}_{(N+2)k} h_k]  \sin \xi ) 
		\right. \right. \\ \left. \left.
+  {i} \bar{\cal F}'_{i} \gamma_5 {\cal F}'_j (- [{\cal O}_{(N+1)k} h_k]  \sin \xi + [{\cal O}_{(N+2)k} h_k]  \cos \xi ) \right] \frac{v_a + [{\cal O}_{ak} h_k]}{w\sqrt{2}} \right].
\end{split}
\end{eqnarray}
Thus, to study an specific case we consider that the most relevant mixing occurs between $\phi_0^b$ and $p_1$, then,
	\begin{eqnarray}  \begin{split}
		-{\cal L}_Y^{f} \approx \sum_{i,j, f} \left \lbrace m_i^f \delta_{ij}
		\bar{F}'_{i} f'_j  + \frac{\widetilde{\bf M}^{f,b}_{ij}}{v_b} \bar{F}'_{i} f'_j  \left( c_{\gamma} h + s_{\gamma} h_{N+2} \right) +  {\textrm{H.c.}}
		 \right. \\
		 +  \sum_{a} r_s^a \widetilde{\bf Z}_{ij}^{f,a} \left[  \bar{\cal F}'_{i} {\cal F}'_j  ( h_{N+1}  \cos \xi +  \left( -s_{\gamma} h + c_{\gamma} h_{N+2} \right)   \sin \xi )
	 \right.\\ \left. + i \bar{\cal F}'_{i} \gamma_5 {\cal F}'_j  (- h_{N+1}  \sin \xi + \left( -s_{\gamma} h + c_{\gamma} h_{N+2} \right)  \cos \xi ) \right] 
		 \Bigg \rbrace, \end{split}
	\end{eqnarray}	
where we have denoted by $h \equiv h_b$ the lightest state and corresponding to the SM-like Higgs,
$r_s^a= v_a/(\sqrt{2}w)$ and where we have taken the limit $w \gg v_a$.

\section{The minimal model with LFV Higgs interactions and CPV: SM Higgs plus one FN singlet\label{Sec:Model}}

Thus, in order to perform our phenomenological study, it is enough to take the simplest picture for the scalar sector, 
assuming  only the SM Higgs doublet, $\Phi$, plus the FN singlet field, {$S_F$}.
Since we are interested in the possibility of having a light SM-like Higgs boson with 
sizeable LFV interactions,  we shall assume that CP is violated in the scalar sector. 
The resulting  mass eigenstates are identified as the SM-like Higgs and the flavons, and one of these flavons 
tends to have a mass of order of the EW scale.
Then, we will use the mixing between the real component of the Higgs doublet and the imaginary component of the flavon 
singlet, to transmit the sizeable LFV interactions to the Higgs boson, which are being tested currently at the LHC.

The scalar fields are then written as follows,
\begin{subequations}
\begin{eqnarray}
 \Phi& =& \left( \begin{array}{c} G^+ \\ \frac{1}{\sqrt{2}} \left( v + \phi^0 + i G_z \right)
\end{array} \right), \label{dec_doublets} \\ \nonumber \\
 S_F &= &\frac{1}{\sqrt{2}} ( w e^{i\xi}  + s_1 + i p_1 ), \label{dec_singlet}
\end{eqnarray}
\end{subequations}
where $v$ denotes the SM vev, while $ u = w e^{i\xi}$  denotes the complex vev of the FN singlet. As we said earlier, the neutral component of the doublet mixes with the imaginary component of the FN singlet. For that,
we shall also use $u_1= w \cos \xi$ and $u_2= w \sin \xi$ for the real and imaginary components of the FN singlet vev, respectively.

\subsection{The Higgs potential\label{subsec:higgspotential}}
The Higgs potential now involves two new parts besides the SM one,
\begin{equation}
 V=  V_{\phi}  +  V_S  +  V_{S\phi}
\end{equation}
where,
\begin{subequations}
\begin{eqnarray}
V_{\phi} &=& -\frac{1}{2}{m_{1}^2} \Phi^\dagger \Phi  
+ \frac{1}{2}\lambda_1 \left( \Phi^\dagger \Phi \right)^2,\\
V_S &=& -\frac{m^2_{s}}{2} S^*_F S_F -\frac{\mu^2_{s}}{2} (S^{*2}_F + S^2_F) + \lambda_{s} (S^*_F S_F)^2
      + \lambda_{s1} S^*_F S_F (S^{*2}_F + S^2_F)   \\
    & &  + \lambda_{s2} (S^{*4}_F + S^4_F) + w \tilde{\lambda}_{sa} (S^{*3}_F + S^3_F) + w \tilde{\lambda}_{sb} (S^*_F S_F) (S^{*}_F + S_F),
 \nonumber\\
V_{S\phi} &=& \lambda_{11} (\Phi^\dagger\Phi)(S^*_F S_F) + {\lambda}_{12} (\Phi^\dagger\Phi) (S^{*2}_F + S^2_F)
                 + w \tilde{\lambda}_{sc} (\Phi^\dagger\Phi) (S^{*}_F + S_F).
                 \end{eqnarray}
\label{potIDM1S}
\end{subequations}
$w$ denotes a dimensional mass scale that allows to write the trilinear terms in terms of such scale
and dimensionless coefficients $\tilde{\lambda}_{si}$. The parameters $m_{1, s1, s2}^2$, $\lambda_{s, s1, s2, 11,12}$ and $\tilde{\lambda}_{sa, sb, sc}$ are all real. Thus, the Higgs potential in Eq.~(\ref{potIDM1S}) depends on eleven real parameters, a total of twelve  degrees of freedom, but how many are physical? To find this out let us first study the minimization conditions.

The minimization conditions read,
\begin{subequations}
\label{eq:minimizationconditions}
\begin{eqnarray}   
  m^2_1  &=&  v^2  \lambda_1 + u_1^2 (\lambda_{11} + 2\lambda_{12}) + u_2^2 (\lambda_{11} - 2\lambda_{12}) +2\sqrt{2} \, u_1 w \tilde{\lambda}_{sc},  \\ 
  m^2_{s}  &=& v^2 \lambda_{11} + 2 u_1^2 \lambda_{s12}^+ +2 u_2^2 \lambda_{s12}^- \\  \nonumber
 &&  + \frac{\sqrt{2}\, w}{2 u_1} \left(v^2 \tilde{\lambda}_{sc} - u_1^2 \left(3\tilde{\lambda}_{sa} - 5\tilde{\lambda}_{sb} \right) + u_2^2 \left(3\tilde{\lambda}_{sa} - \tilde{\lambda}_{sb} \right)  \right) , \\ \nonumber
{\mu}^2_{s} &=&   v^2 \lambda_{12} +  u_1^2 \left(\lambda_{s1} + 4 \lambda_{s2}\right) + u_2^2 \left(\lambda_{s1} - 4\lambda_{s2} \right) \\ \label{mincondition}
 &&  + \frac{\sqrt{2}\, w}{4 u_1} \left(v^2 \tilde{\lambda}_{sc} + u_1^2 \left(9\tilde{\lambda}_{sa} + \tilde{\lambda}_{sb} \right) - u_2^2 \left(3\tilde{\lambda}_{sa} - \tilde{\lambda}_{sb} \right)  \right)  ,
\end{eqnarray}
 \end{subequations}
with  $\lambda_{s12}^{\pm} = \lambda_s{\pm}\lambda_{s1} - 2 \lambda_{s2}$. 

 We now consider the mixture between the neutral component of the Higgs field with the other two degrees of freedom of the complex singlet.  The mass matrix in this basis ($\phi^0, s_1, p_1$), has as eigenvalues the neutral Higgs masses. This matrix, a $3 \times 3$ symmetric one, is here denoted as $M^2_s$. A simplification can be achieved by using the minimization conditions (\ref{eq:minimizationconditions}); as a consequence the entries of this matrix are then given as,
\be
\label{eq:massmatrix}
\begin{array}{rcl}
M^2_s(1,1) & = & v^2 \lambda_{1} ,   \\
M^2_s(1,2) & = & v\left(u_1(\lambda_{11} + 2 \lambda_{12}) + \sqrt{2} \, w\tilde{\lambda}_{sc}\right),    \\
M^2_s(1,3) & = & v u_2  (\lambda_{11} - 2 \lambda_{12}) ,   \\
M^2_s(2,2) & = & 2 u_1^2\left(\lambda_{s} + 2(\lambda_{s1}+\lambda_{s2})\right) \\
                   &     & +\displaystyle\frac{\sqrt{2}\, w}{2u_1} \left( 3u_1^2(\tilde{\lambda}_{sa} +  \tilde{\lambda}_{sb}) + u_2^2(3\tilde{\lambda}_{sa} -  \tilde{\lambda}_{sb}) - v^2 \tilde{\lambda}_{sc} \right) ,    \\
M^2_s(2,3) & = & u_2\left(2u_1(\lambda_s-6\lambda_{s2}) - \sqrt{2}\, w (3\tilde{\lambda}_{sa} -  \tilde{\lambda}_{sb})  \right),   \\
M^2_s(3,3) & = & 2u_2^2 (\lambda_s - 2\lambda_{s1} + 2 \lambda_{s2}).
\end{array}
\ee

Notice that there is mixing between the real and imaginary components of the scalar fields, which in principle point to the appearance of CP violation in the scalar sector.

The symmetric matrix $M^2_s$ is diagonalized by the orthogonal matrix ${\cal O}$,
\be
\label{diagonal}
{\cal O}^T M^2_s {\cal O} = \mathrm{diag}(m_{h_1}^2,m_{h_2}^2,m_{h_3}^2) ,
\ee
where
\begin{eqnarray}
{\cal O} &=& T_1 T_2 T_3 ,
\end{eqnarray}
and
\be
\label{eq:tmatrix}
T_1 = \left(
\begin{array}{ccc}
 c_{\alpha_1} & s_{\alpha_1} & 0 \\
 -s_{\alpha_1} & c_{\alpha_1} & 0 \\
0 & 0 & 1
\end{array} \right),
\quad
T_2 = \left(
\begin{array}{ccc}
 c_{\alpha_2} & 0 & s_{\alpha_2}  \\
 0 & 1 & 0 \\
 -s_{\alpha_2}& 0 & c_{\alpha_2}  
\end{array} \right),
\quad
T_3 = \left(
\begin{array}{ccc}
1 & 0 & 0 \\
0 & c_{\alpha_3}&  s_{\alpha_3}  \\
0 & -s_{\alpha_3}&  c_{\alpha_3}  
\end{array} \right).
\ee
Obtaining after substitution,
\be
\label{eq:ortogonalmatrix}
\begin{array}{rcl}
{\cal O}
&=&  \left(
\begin{array}{ccc}
c_{\alpha_1}c_{\alpha_2} & c_{\alpha_3} s_{\alpha_1} - c_{\alpha_1} s_{\alpha_2} s_{\alpha_3} & c_{\alpha_1} c_{\alpha_3} s_{\alpha_2}+ s_{\alpha_1} s_{\alpha_3} \\
-c_{\alpha_2}s_{\alpha_1} & c_{\alpha_1} s_{\alpha_3} + s_{\alpha_1} s_{\alpha_2} s_{\alpha_3} & c_{\alpha_1} s_{\alpha_3} - c_{\alpha_3}+ s_{\alpha_1} s_{\alpha_2} \\
-s_{\alpha_2} & -c_{\alpha_2} s_{\alpha_3}  & c_{\alpha_2} c_{\alpha_3} 
\end{array} \right) .
\end{array}
\ee
%
\subsection{Analysis of the Higgs masses and mixing\label{subsec:analhiggsmassmix}}
For the sake of simplification, let us consider the particular case,
\be
\lambda_{12} \thicksim \lambda_{11} \thicksim  \lambda_{s2} \thicksim \lambda_{s1},
\label{eq:particularcase}
\ee
which could be motivated by the use of some symmetrical argument but that we do not explore here.

Then, the scalar mass matrix takes the form,
{
\begin{small}
\begin{eqnarray} \label{mssimplificada}
{\bf M}_s^2 = \begin{pmatrix}
 v^2 \lambda_{1}    &  v 3u_1\lambda_{11}   & -v u_2 \lambda_{11}    \\
 v 3u_1\lambda_{11}  &  2u_1^2(\lambda_s + 4\lambda_{s1}) +\displaystyle\frac{ \sqrt{2}\, w}{2u_1}\left(2(3 u_1^2 +  u_2^2) \tilde\lambda_{sa}  \right) & u_2(2u_1(\lambda_s-6\lambda_{s1})-2\sqrt{2} \, w \tilde\lambda_{sa})  \\
 -v u_2 \lambda_{11} & u_2(2u_1(\lambda_s-6\lambda_{s1})-2\sqrt{2} \, w \tilde\lambda_{sa})  & 2u_2^2\lambda_s
\end{pmatrix}.
\end{eqnarray} 
\end{small}}

From Eq.~(\ref{diagonal}) and Eq.~(\ref{mssimplificada}), we have got that 
$\lambda_{1}$, $\lambda_{s}$, $\lambda_{s1}$, $\lambda_{11}$, $\tilde\lambda_{sa}$, and $\tilde\lambda_{sc}$ 
are determined by $m_{h_1,h_2,h_3}^2, v, w, u_1$, and $u_2$.

Through a scanning of the space of parameters we see that when,
$$
M^2_{h_1}  \ll M^2_{h_3} \ll M^2_{h_2},
$$
then, we shall consider the following spectrum,
\be
\begin{array}{c}
123 \ \mathrm{GeV} \leq M_{h_1} \leq 126 \ \mathrm{GeV}, \\
500 \ \mathrm{GeV} \leq M_{h_2} \leq 1000 \ \mathrm{GeV}, \\
150 \ \mathrm{GeV} \leq M_{h_3} \leq 500 \ \mathrm{GeV}. \\
\end{array}
\ee
Then, one can take:  $\lambda_1=0.125$, as in the SM, which is a good approximation. For the mixing we shall consider that $0 \leq \alpha_2 \leq \pi$, while:
\be
\alpha_1 \leq \pi/32, \quad \qquad \text{and} \qquad \quad \alpha_3 \leq \pi/32 \, ,
\ee 

For the CPV phase $\xi$ we shall take,
\be
 0 \leq \xi \leq 2 \pi,
\ee
and fixing $v=246$ GeV, we shall use the ratio:
\be
r_s = \frac{v}{ \sqrt{2} w} , 
\ee
with $0.5 \leq w \leq 10$ TeV.
\begin{figure}[t]
 \begin{center}
  \begin{tabular}{cc}
\includegraphics[width=7.7cm, height=5.3cm]{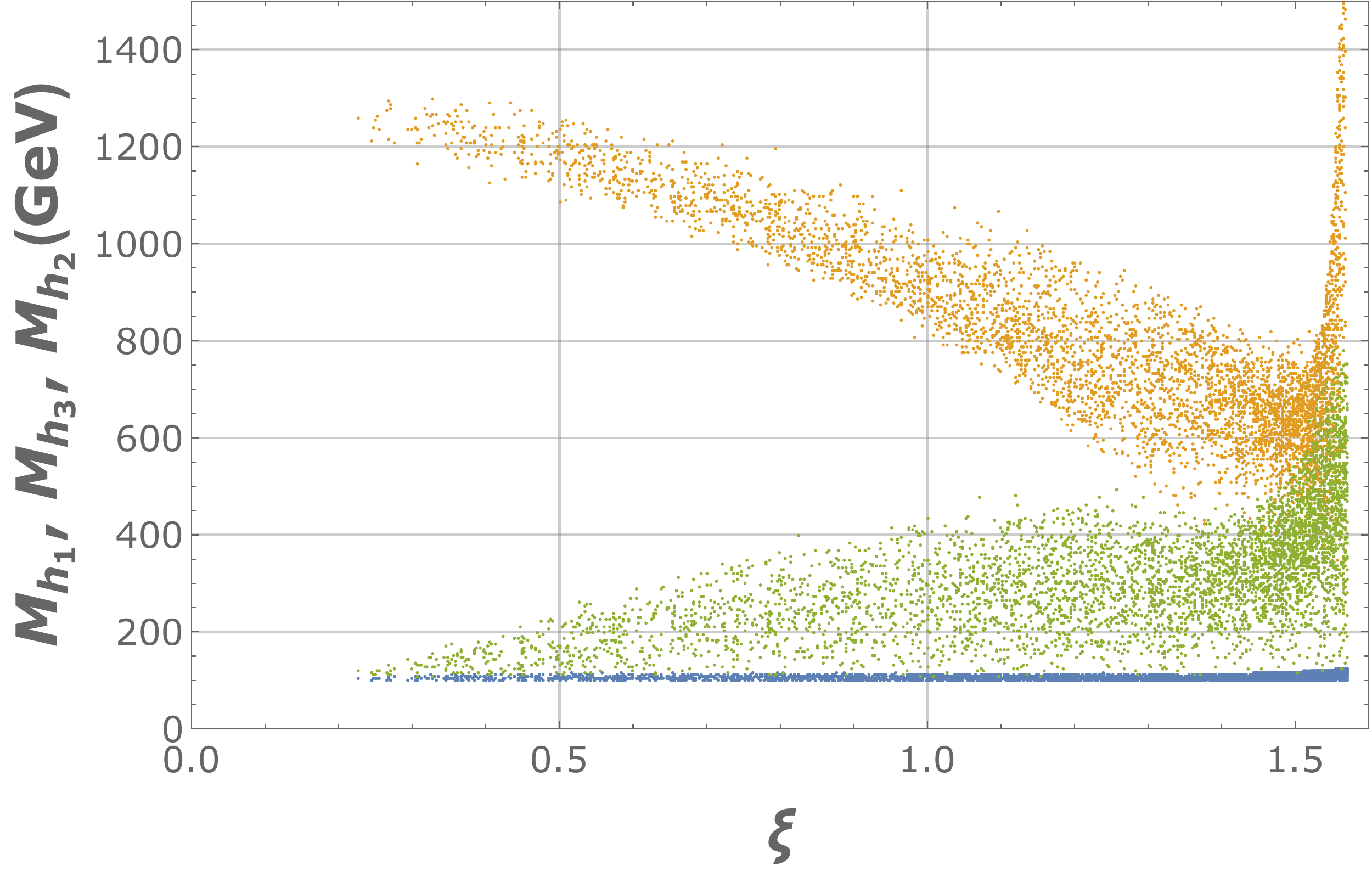}  
\includegraphics[width=7.7cm, height=5.3cm]{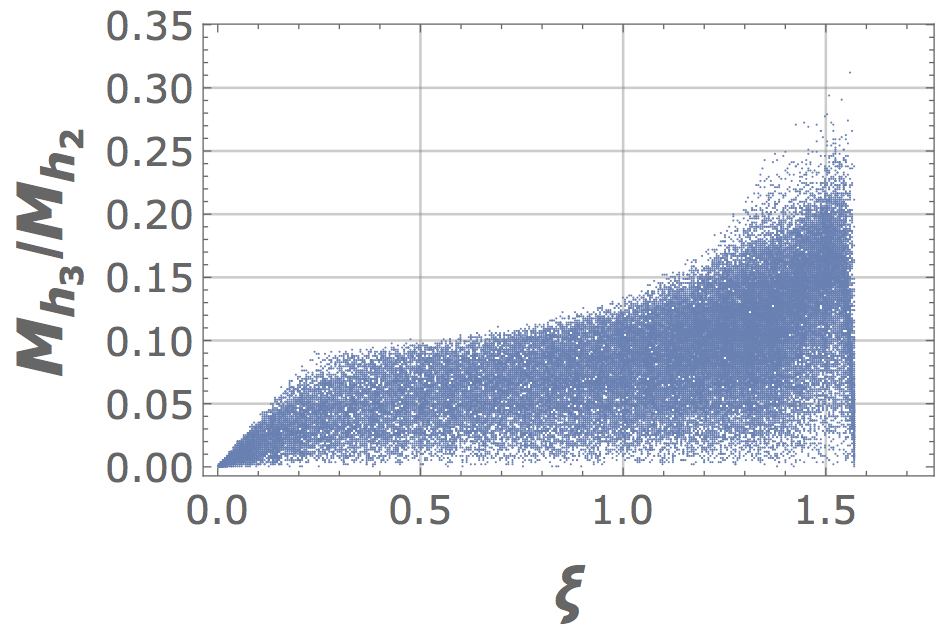}
        \end{tabular}
  \caption{Neutral Higgs masses. Left panel shows the different mass values $M_{h_1}$ (blue color), $M_{h_2}$ (green color), and $M_{h_3}$ (orange color) as function of the CPV phase parameter $\xi$ in the allowed range $0<\xi<1.5\,{\textrm{rad}}$. The lightest mass state is the one assigned equivalent to the SM Higgs state, $M_{h_1} \sim M_{h_{SM}}$. On the other hand, the right panel plots the ratio among the other two neutral mass states, $M_{h_3}/M_{h_2}$, as function of the same CPV phase parameter. For both plots, we have considered $w=0.5$ TeV.}
   \label{Fig:masses1}
 \end{center}
\end{figure}

In Fig.~\ref{Fig:masses1}, we show the mass spectra for the three neutral scalar states. There we have assumed $w=0.5$ TeV. The behavior shown around the CPV phase value, $\xi = \pi/2$, describes the situation when the mass matrix elements diverge. Due to this we have restrained ourselves to the region, $0 \leq \xi <\pi/2$.

\section{The FN mechanism  and the Yukawa matrices\label{Sec:Matrices}}
Not only we are extending in a simple way the field content of the SM but 
we are also incorporating the so called FN mechanism \cite{Froggatt:1978nt}, by which we are considering a theory with a mechanism to produce hierarchical Yukawa couplings (which is later translated into hierarchical fermion masses). Our approach in this paper consists in exploiting the general features FN models have. The following calculations are shared in general by all those models having a single flavon field. Furthermore, {in the second part of this section}, we also consider a general property Yukawa couplings have and we show how it is enough to study the coupling between the tau and the {SM-like} Higgs field. 

\subsection{Flavons and the Yukawa Lagrangian\label{subsec:fylag}}

%
As we are considering spontaneous CP violation coming from the flavon field, we have assigned it a complex vev,
\be
u= w e^{i \xi}. 
\ee
It turns out that the angle $\xi$, mixes the scalar and pseudoscalar CP states of the Higgs and flavon fields. In fact, it has been already called the 
scalar-pseudoscalar Higgs mixing angle. Now, the general Yukawa Lagrangian (Eq.~(\ref{Eq16})) becomes for this simplest model, the following,
\be
-{\cal L}_Y^\ell = \sum_{i,j} \left[ {\bf M}^\ell_{ij} 
\bar{E}_{i} e_j  +  \frac{{\bf M}^\ell_{ij}}{v} \bar{E}_{i} e_j \phi_0 
+ {\bf Z}_{ij}^\ell e^{-i\xi }\frac{s_1+ip_1}{w} \bar{E}_{i} e_j \frac{v + \phi_0}{\sqrt{2}}+  {\textrm{H.c.}} \right],
\ee
where ${\bf M}_{ij}^\ell = \frac{v}{\sqrt{2}} Y_{ij}^\ell$, ${\bf Z}_{ij}^\ell = \kappa_{ij} {Y}_{ij}^\ell$, and we have assumed the unitary gauge $G_z \rightarrow 0$.

\paragraph*{Digression:}
Let us consider that ${\bf Z}_{ij}^\ell = {\bf Z}_{ij}^{\ell^{\dagger}}$\footnote{ At this point, we have partially lost the proposed generality of using FN model 
features by winning an easiness in our calculations. Notice that to achieve such an Hermitian structure one could think on models which have a symmetrical position of their powers along the off-diagonal elements, that is, $\kappa_{ij} = \kappa_{ji}$  
($i \neq j$).}. This means, when combined with the Hermitian conjugate, that the following term acquires the form,
\be
\begin{array}{rcl}
{\bf Z}_{ij}^\ell e^{-i\xi } \left( \frac{s_1+i p_1}{w} \right) \bar{E}_{i} e_j \left( \frac{v + \phi_0}{\sqrt{2}} \right)+ {\textrm{H.c.}}, 
&=&{\bf Z}_{ij}^\ell e^{-i\xi } \left( \frac{s_1 +ip_1}{w} \right) \bar{\ell}_{i} P_R \ell_j \left( \frac{v + \phi_0}{\sqrt{2}} \right)
+ {\bf Z}_{ij}^\ell{}^\dagger e^{i\xi } \left( \frac{s_1-ip_1}{w}\right) \bar{\ell}_{i} P_L \ell_j \left( \frac{v + \phi_0}{\sqrt{2}} \right), \\
&=&{\bf Z}_{ij}^\ell \left[ \bar{\ell}_{i} \ell_j (s_1 \cos \xi + p_1 \sin \xi )
+  {i} \bar{\ell}_{i} \gamma_5 \ell_j (- s_1 \sin \xi + p_1 \cos \xi ) \right] \frac{v + \phi_0}{w\sqrt{2}},
\end{array}
\ee
where $P_{L} \ell = E$, $P_R \ell = e$, and $P_{L(R)} = \frac{1\mp \gamma_5}{2}$.
	
Coming back to the Lagrangian. In the limit $w \gg v$ we are left with,
\be
-{\cal L}_Y^\ell = \sum_{i,j} \Big \lbrace {\bf M}^\ell_{ij} 
\bar{E}_{i} e_j  +  \frac{{\bf M}^\ell_{ij}}{v} \bar{E}_{i} e_j \phi_0 +  {\textrm{H.c.}}
+ r_s {\bf Z}_{ij}^\ell \left[  \bar{\ell}_{i} \ell_j  (s_1 \cos \xi + p_1 \sin \xi ) i \bar{\ell}_{i} \gamma_5 \ell_j  (- s_1 \sin \xi + p_1 \cos \xi ) \right] 
\Big  \rbrace,
\ee	
where we have denoted $r_s \equiv \frac{v}{w\sqrt{2}}$.
	
Now, we introduce the following mass eigenstates, $h_j$, from the scalar sector,
\be
\begin{pmatrix}
			\phi_0  \\
			s_1 \\
			p_1 
		\end{pmatrix}
		= \begin{pmatrix}
			{\cal O}_{1j} h_j \\
			{\cal O}_{2j} h_j \\
			{\cal O}_{3j} h_j
\end{pmatrix}.
\ee
We are then allowed to substitute them and obtain
\be
\begin{array} {rcl}
-{\cal L}_Y^\ell &=& \sum_{i,j} \left \lbrace {\bf M}^\ell_{ij} 
\bar{E}_{i} e_j  +  \frac{{\bf M}^\ell_{ij}}{v} \bar{E}_{i} e_j [{\cal O}_{1k} h_k]  +  {\textrm{H.c.}}
 \right. \\
&+& r_s {\bf Z}_{ij}^\ell \left[  \bar{\ell}_{i} \ell_j  ([{\cal O}_{2k} h_k]  \cos \xi + [{\cal O}_{3k} h_k]  \sin \xi )
+  i \bar{\ell}_{i} \gamma_5 \ell_j  (- [{\cal O}_{2k} h_k]  \sin \xi + [{\cal O}_{3k} h_k]  \cos \xi ) \right] 
		 \Big \rbrace, 
\end{array}
\ee	

To study an specific case we consider that the most relevant mixing occurs between $\phi_0$ and $p_1$, then,
\begin{eqnarray}
		{\cal O} \simeq T_2 = \begin{pmatrix}
			c_{\alpha_2} & 0 & s_{\alpha_2} \\
			0 &  1 & 0 \\
			-s_{\alpha_2} & 0 & c_{\alpha_2}
		\end{pmatrix}.
\end{eqnarray}
Afterwards, it is easy to see that,
\begin{eqnarray} \begin{split}
		-{\cal L}_Y^\ell = \sum_{i,j} \left \lbrace {\bf M}^\ell_{ij} 
		\bar{E}_{i} e_j  +  \frac{{\bf M}^\ell_{ij}}{v} \bar{E}_{i} e_j  \left( c_{\alpha_2} h + s_{\alpha_2} h_3 \right) +  h.c.
		 \right. \\
		 + r_s {\bf Z}_{ij}^\ell \left[  \bar{\ell}_{i} \ell_j  ( h_2  \cos \xi +  \left( -s_{\alpha_2} h + c_{\alpha_2} h_3 \right)   \sin \xi )
	+  i \bar{\ell}_{i} \gamma_5 \ell_j  (- h_2  \sin \xi + \left( -s_{\alpha_2} h + c_{\alpha_2} h_3 \right)  \cos \xi ) \right] 
		 \Bigg \rbrace, \end{split}
\end{eqnarray}	
where we have denoted by $h \equiv h_1$ the lightest state and corresponding to the SM-like Higgs. 
	
After moving the lepton fields to the mass basis we get, after some reordering,
\begin{eqnarray} \begin{split}
-{\cal L}_Y^\ell = \sum_{i,j} \left[ m_i^\ell \delta_{ij} 
\bar{\ell}'_{i} \ell'_j  + \bar{\ell}'_{i}  \left(c_{\alpha_2} \frac{m_i^\ell \delta_{ij}}{v}
-s_{\alpha_2} r_s \widetilde{\bf Z}_{ij}^\ell (\sin\xi + i \gamma_5 \cos\xi)
\right) 	\ell'_j h \right. \\
 + r_s \widetilde{\bf Z}_{ij}^\ell  \bar{\ell}'_{i} (\cos\xi - i \gamma_5 \sin \xi) \ell'_j h_2  \left. 
 +\bar{\ell}'_{i}  \left(s_{\alpha_2} \frac{m_i^\ell \delta_{ij}}{v}
+ic_{\alpha_2} r_s \widetilde{\bf Z}_{ij}^\ell  (\sin\xi + i \gamma_5 \cos \xi)
\right) \ell'_j h_3 \right].
\end{split}
\end{eqnarray}

Our neutral SM-like Higgs boson belongs to a CP mixture state as it
couples to both the scalar and the pseudoscalar fermion currents. {  Similar couplings can be obtained for the couplings with d- and u-type quarks with the replacements $m_i^\ell \rightarrow m_i^d, \; m_i^u$ and ${\bf Z}_{ij}^\ell \rightarrow {\bf Z}_{ij}^q$. Thus the FC fermionic couplings of the lightest Higgs state $h$ will be of the form,} 
{  
	\begin{eqnarray}
		( h f_i \bar{f}_i ) = i( a_i - ib_i \gamma_5)
	\end{eqnarray}
	where,
	\begin{eqnarray}
		a_i = c\alpha_2 \frac{m_i}{v} - s\alpha_2 r_s \widetilde{\bf Z}_{ii}^f \sin \xi ,\quad \quad
		b_i = -s\alpha_2 r_s \widetilde{\bf Z}_{ii}^f \cos \xi .
	\end{eqnarray}
	While the FV ones,
	\begin{eqnarray}
		( h f_i \bar{f}_j ) = i( c_{ij} - i d_{ij} \gamma_5)
	\end{eqnarray}
	with,
	\begin{eqnarray}
		c_{ij} = -	 s\alpha_2 r_s \widetilde{\bf Z}_{ij}^f \sin \xi , \quad\quad \quad
		d_{ij} = - s\alpha_2 r_s \widetilde{\bf Z}_{ij}^f \cos \xi .
	\end{eqnarray}
}
%

\subsection{The two-family approximation\label{subsec:subsec:twofamilyap}}
Now, in order to continue we need to know or at least have a fair estimation of
the $2-3$ sector in $\widetilde{\bf Z}^\ell$. 
For that, we do the following. Before deciding on any particular
model, let us see how much we can gain by just studying shared features inside most
leptonic FN models. For example, 
generically speaking, in these models the ratio between the $\mu-\tau$ sector Yukawa couplings is approximately given as \cite{Chen:2011sb,Dery:2014kxa,Huitu:2016pwk},
\begin{eqnarray}
	\Bigg| \frac{Y_{23}^\ell}{Y_{33}^\ell } \Bigg| \sim \lambda^2, 
\end{eqnarray} 
while the next one changes from model to model,
\begin{eqnarray}
	\Bigg| \frac{Y_{22}^\ell}{Y_{23}^\ell } \Bigg| \sim \lambda^{n},
\end{eqnarray}
with $n= 0,1,2,\dots,$ and for last, the Cheng-Sher ansatz gives a good limit for the off-diagonal elements, 
which in general are not symmetrical, and should fulfill the inequality~\cite{Cheng:1987rs},
\begin{eqnarray}
	| {Y_{23}^\ell}{Y_{32}^\ell } | \leq \frac{m_\mu m_\tau}{v^2}.
\end{eqnarray}
{This ansatz is justified by considering that in order to produce hierarchical masses ($m_\tau \gg m_\mu$) one needs to constrain the off-diagonal entries to be sufficiently small compared to the $\tau \tau$ matrix element.}

So we see how we can already suggest, by the previous arguments, the $2\times 2$ matrix,
\begin{eqnarray}
	{\bf y}_{\mu \tau} \simeq \begin{pmatrix}
		\lambda^{n+2} Y_{33}^\ell  & \lambda^2 Y_{33}^\ell \\
			Y_{32}^\ell & Y_{33}^\ell
	\end{pmatrix}.
\end{eqnarray}

In fact, we do not need to know $Y_{32}^\ell$. As the matrix which is diagonalized is the left Hermitian product, 
\begin{eqnarray}
	{\bf y}_{\mu \tau} {\bf y}_{\mu \tau}^\dagger \approx |Y_{33}^\ell |^2
	\begin{pmatrix}
			\lambda^4(1 + \lambda^{2n}) & \lambda^2 \\
			\lambda^2 & 1	
	\end{pmatrix},
\end{eqnarray}
where for this approximation could be enough to consider that $|Y^\ell_{32}/Y_{33}^\ell | \sim \lambda$.
Now, bear in mind that all of this is being done inside the limit where $m_e \rightarrow 0 $, 
as the electron mass can be safely neglected. The importance of these approximations is that we can actually express the angle,
which helps to diagonalize the 2-3 submatrix, in
terms of the mass ratio,
\begin{eqnarray}
	\theta \simeq \begin{cases}
	\frac{m_\mu}{m_\tau} & n = 0 \\
	 (\frac{m_\mu}{m_\tau})^{1/n} & n > 1
	 \end{cases}.
\end{eqnarray}

It is a straightforward calculation to show that,
\begin{eqnarray}
	|Y_{33}^\ell | \simeq \frac{m_\tau - m_\mu}{v},
\end{eqnarray} 
and also, we can identify \cite{Ibarra:2003xp},
\begin{eqnarray}
	\lambda^2 \approx \frac{m_\mu}{m_\tau} \approx 0.06,
\end{eqnarray}
providing by this a Cabibbo-like value for the lepton sector, $\lambda \approx 0.24$.

After all this digression, we can now compute what range of values the matrix element $Z_{33}^\ell$ should have. The $2\times 2$ submatrix has acquired the form,
\begin{eqnarray}
	{\bf z}_{\mu \tau} \simeq \begin{pmatrix}
		(n+j+2) \lambda^{n+2} Y_{33}^\ell  & (j+2) \lambda^2 Y_{33}^\ell \\
			(j+2) \lambda^2 Y_{33}^\ell & j Y_{33}^\ell
	\end{pmatrix},
\end{eqnarray}
where Hermiticity implied taking the particular scenario where $Y_{32}^\ell = (j+2) \lambda^2 Y_{33}^\ell$ and $j$ represents the power of $\lambda$ in $Y_{33}^\ell$. 

Finally, in the mass basis, we obtain,
\begin{eqnarray}
	\widetilde{\bf z}_{\mu \tau} \simeq \frac{m_\tau - m_\mu}{v} 
	\begin{cases}
		\begin{pmatrix}
			(j+2)\lambda^3 & 2(j+1)\lambda^2 \\
			2(j+1)\lambda^2 & j
		\end{pmatrix}  & n= 0, \\
		{} \\
		\begin{pmatrix}
			j\lambda^2 & j\lambda \\
			j\lambda & j
		\end{pmatrix} & n= 2,
	\end{cases}
	\label{eq:casesvertex}
\end{eqnarray}
where we have only considered the two most common cases and from which it is interesting to note 
that for the Gatto--Sartori--Tonin-like relation the couplings are less suppressed, for more details about this relation see \cite{Gatto:1968ss} and for a more recent study \cite{Saldana-Salazar:2016hxb}.


\section{Higgs constraints and LFV Higgs decays\label{Sec:Higgs-constraints}}

\subsection{Higgs decays\label{Sec:HiggsDecays}}
Calculating the Higgs decays widths is one of the first steps in order to study the Higgs phenomenology. As the lightest scalar state of our model will be identified with the 125 GeV Higgs boson observed at the LHC, modulo the uncertainties, only certain decay modes will be relevant. 

Within our model there are three scalar mass eigenstates arising from the mixing of real and imaginary components of the doublet and singlet which signal the CP violation. 
The FC/FV couplings with fermions will be of the form,
\begin{eqnarray}
(h f_i \bar{f}_i) = i (a_i - i b_i \gamma_5), \hspace{1 cm} (\text{FC}),\; \\
(h f_i \bar{f}_j) = i (c_{ij} - i d_{ij} \gamma_5), \hspace{1 cm} (\text{FV}).
\end{eqnarray} 
Thus, the decay into FC modes are given by,
\begin{eqnarray}
\Gamma (h_1 \rightarrow f_i \bar{f}_i) = N_c [f(m_i,m_h)] (a_i^2 + b_i^2),
\end{eqnarray}
while the FV ones,
\begin{eqnarray}
	\Gamma (h_1 \rightarrow f_i \bar{f}_j) = N_c [g(m_i,m_j,m_h)] (c_{ij}^2 + d_{ij}^2).
\end{eqnarray}
Notice that the FC modes are sensitive to the CPV phase, $\xi$, due to the interference of the
scalar SM-like coupling ($\propto m_i$) with the flavon-like coupling ($\propto \widetilde{\bf Z}_{ij}^f$).

Next, we have the three-body decays $h \rightarrow WW^*$ and $h\rightarrow ZZ^*$, which
can be written as,
\begin{eqnarray}
	\Gamma (h \rightarrow VV^*) = c^2\alpha_2 \Gamma (\phi_{SM} \rightarrow VV^*),
\end{eqnarray}
where $\Gamma (\phi_{SM} \rightarrow VV^*)$ denotes the decay width of the SM Higgs. 


\subsection{LHC Higgs constraints}
Recent data on Higgs physics coming from the ongoing experiments at the LHC can be employed to derive bounds on the Higgs couplings,
which deviate from the SM in our models. Following ref.~\cite{Giardino:2013bma}, one
has bounds on the parameters $\epsilon_X$, defined as the (small) deviations of the 
Higgs couplings from the SM values, \textit{i.e.} $g_{hXX}= g^{sm}_{hXX} (1 + \epsilon_X)$.
We write our parameters as: $|\eta_X|= 1 + \epsilon_X$. 
The allowed values for fermions are:
$\epsilon_t= -0.21\pm 0.22$, $\epsilon_b= -0.19\pm 0.30$, $\epsilon_{\tau}= 0.00 \pm 0.18$;
while for the W and Z bosons these numbers are: $\epsilon_W= -0.15\pm 0.14$ and $\epsilon_Z= -0.01\pm 0.13$.

In the mass basis, for both the fermions and scalar fields, couplings between the charged fermions and the SM Higgs field {are of the type, $a+ib\gamma_5$. Thus the analysis of ref.~\cite{Giardino:2013bma} is not completely valid for our model. However, we shall focus on the CP-even observables, which are proportional to $(a^2 + b^2)$. These include the analysis of the couplings $hb\bar{b}$, $h\tau \tau$ and we can also use results of ref.~\cite{Giardino:2013bma} for the couplings of $hWW$ and $hZZ$, while the cases $h\gamma \gamma$ and $hgg$ couplings get more complicated. The $ht\bar{t}$ coupling deserves also a separate treatment, as it is derived indirectly from the gluon fusion loop coupling. Thus we shall assume that the bounds on the couplings $hbb$, $h\tau\tau$, $hWW$, and $hZZ$ are still valid, provided that $|\eta| = \sqrt{a^2 + b^2}$ for the fermionic couplings,
where $i = 1,\; 2,\; 3$ and $f= u,\; d,\; e$. Likewise, from the latter we may easily
find $\epsilon_X \equiv |\eta_X| - 1$ with $X$ being the fermion field, $(f,i)$.  For last, couplings between the gauge bosons and the SM Higgs field are changed to,
\begin{eqnarray}
	g_{hVV}^{th} = \cos (\alpha_2) g_{hVV}^{SM}.
\end{eqnarray}

In Fig.~\ref{Fig:Couplings}, we show the deviation coupling $\epsilon_\tau$ as a function of the angle $\alpha_2$  
for three different cases, $w = (0.5, 1.0, 5.0)$ TeV.  
We shall only obtain specific points in parameter space 
which satisfy the LHC bounds. These points will then be used in our analysis of LFV Higgs decays. 
On the other hand, in the same figure, Fig.~\ref{Fig:Couplings} , we show
the behavior of $\epsilon_Z$ and $\epsilon_W$ as function of $\alpha_2$ irrespective of $w$. 

One specific region, in agreement with all data, is: $\alpha_2 < 0.4$.

\begin{figure}[!htbp]
  \begin{tabular}{cc}
  \subfigure[]{ \includegraphics[width=7.7cm, height=5.3cm]{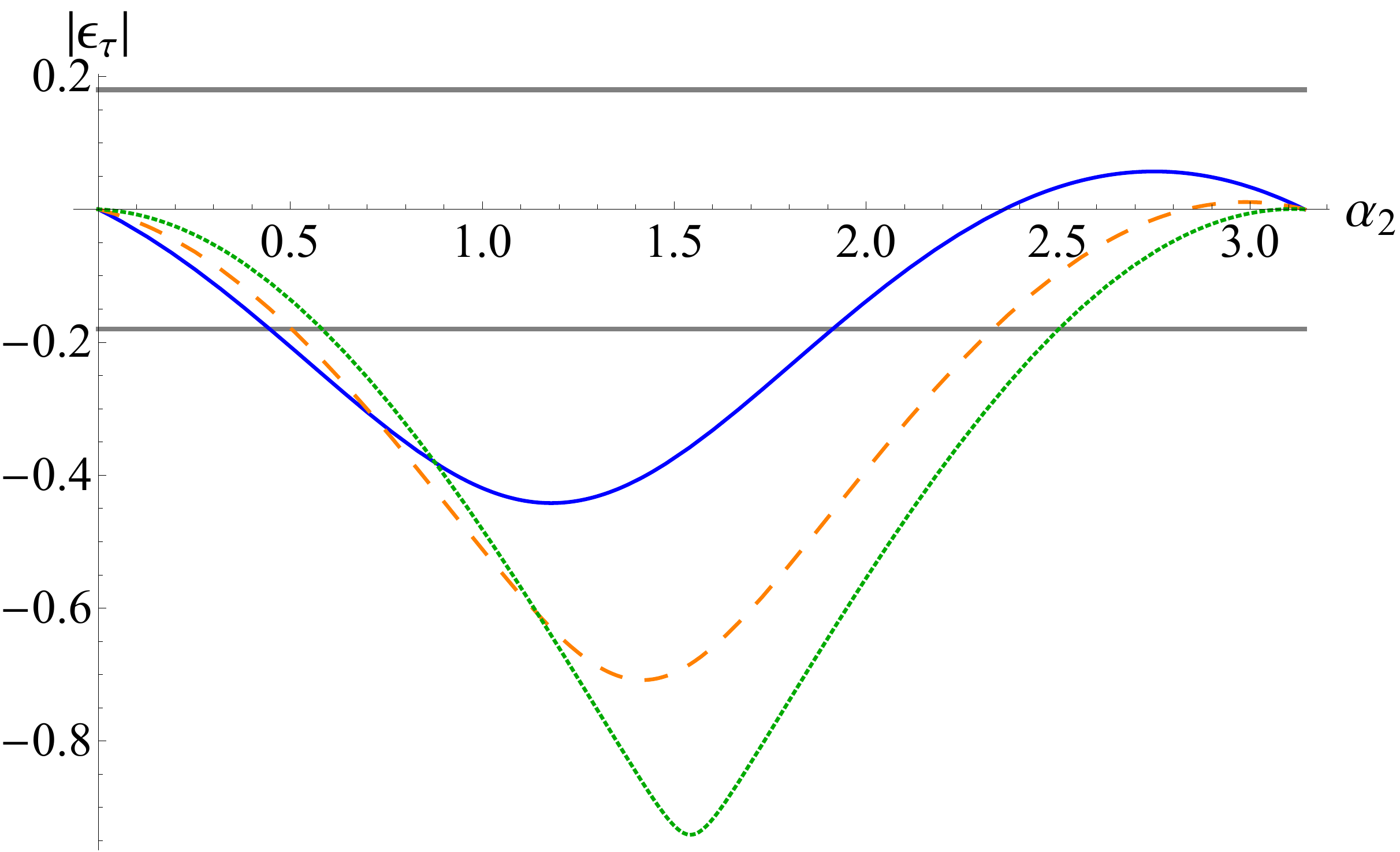}}
  \subfigure[]{ \includegraphics[width=7.7cm, height=5.3cm]{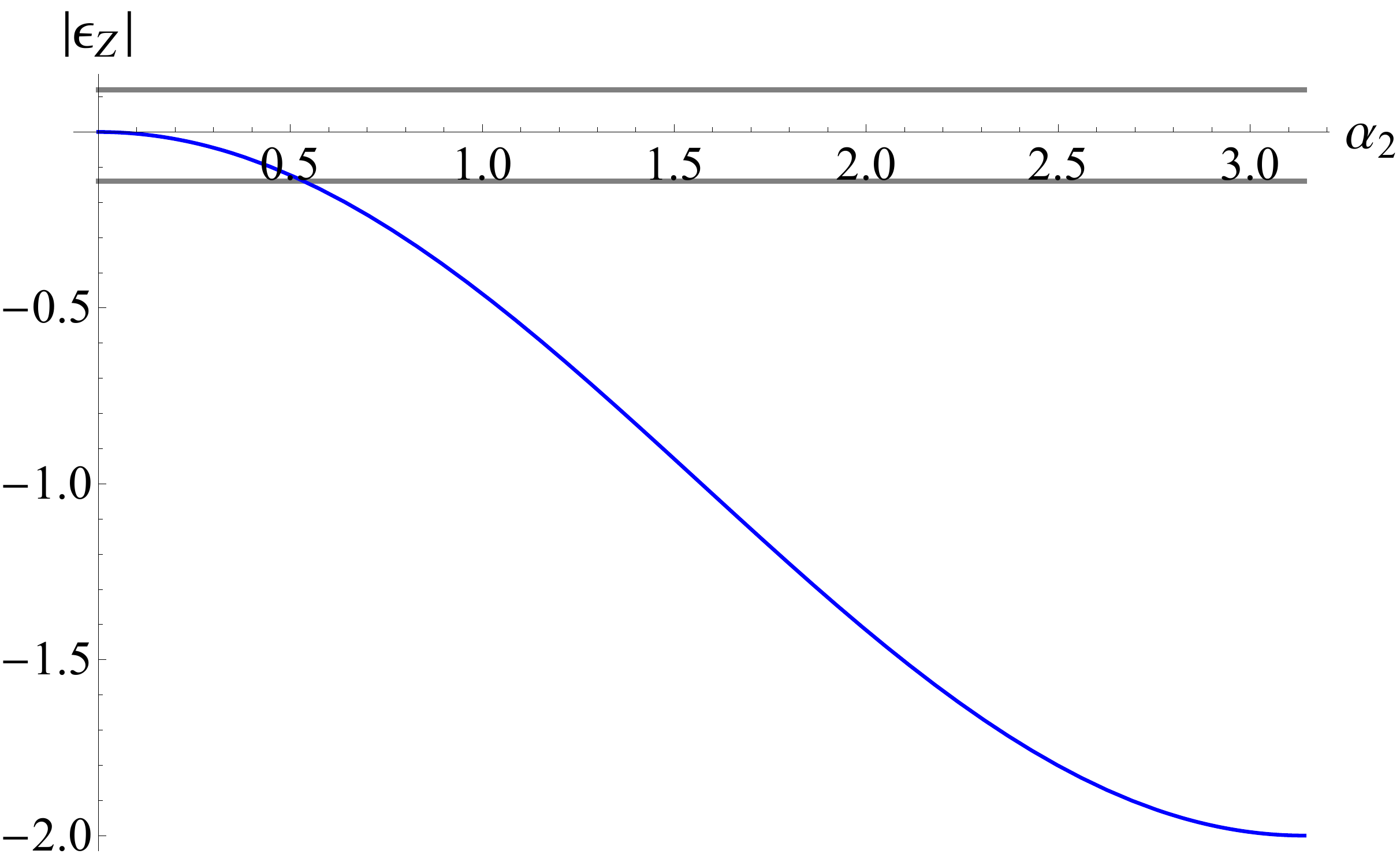}} \\
\subfigure[]{ \includegraphics[width=7.7cm, height=5.3cm]{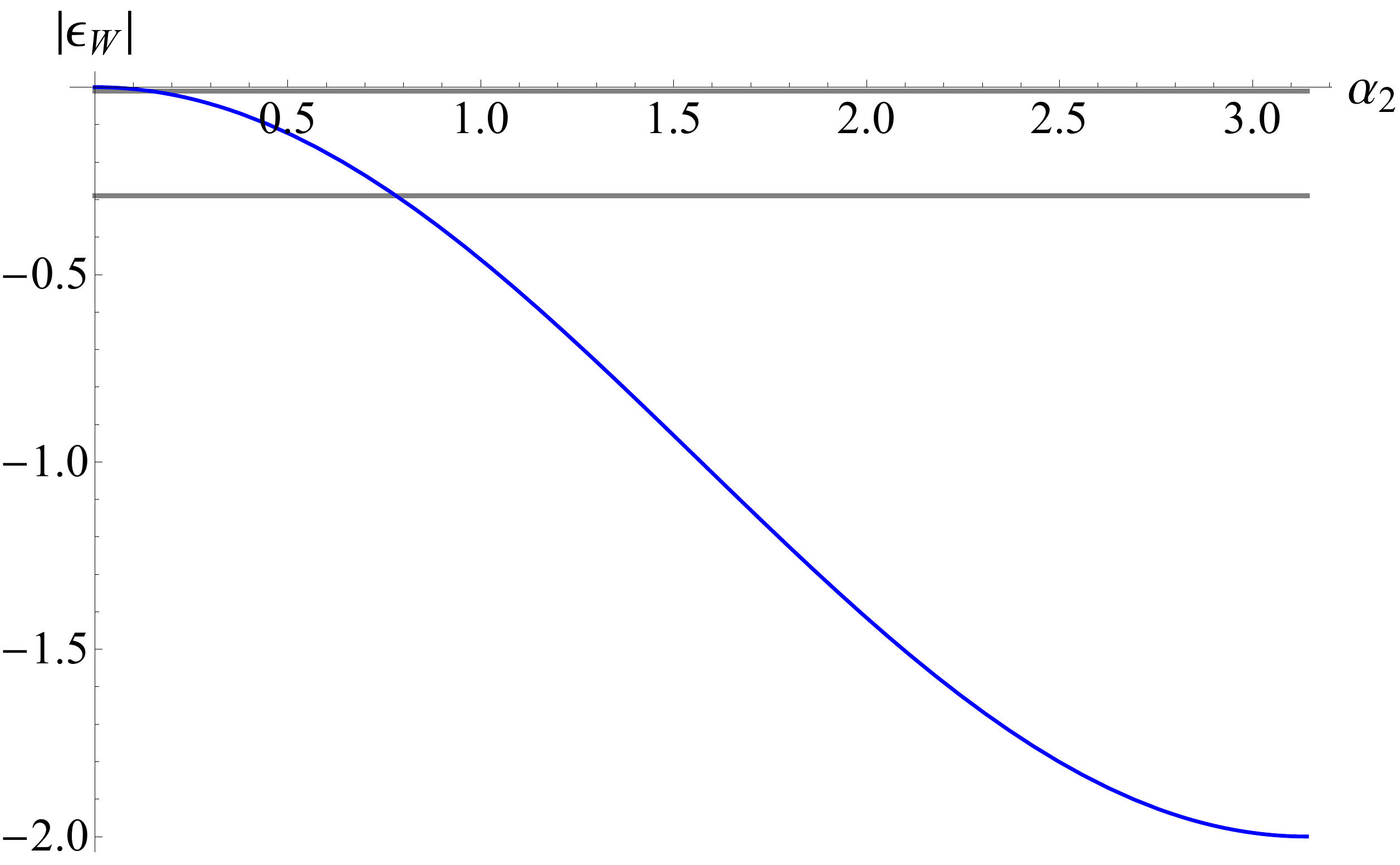}} 
  \end{tabular}
  \caption{The deviations of the Higgs couplings from the SM values are defined by $g_{hXX}= g^{sm}_{hXX} (1 + \epsilon_X)$. 
Here they are shown the small deviations, $\epsilon_\tau$,  $\epsilon_Z$, and  $\epsilon_W$ to the tau lepton, the $Z$ boson, and the $W$ boson, respectively, as function of $\alpha_2$. 
The horizontal lines are the experimental limits on each factor, $\epsilon_{\tau}= 0.00 \pm 0.18$, $\epsilon_Z= -0.01\pm 0.13$, and $\epsilon_W= -0.15\pm 0.14$~\cite{Giardino:2013bma}.
Only in the tau coupling exist a dependence to $w$, thus, the continuous (blue) line,
large dashing (orange) line, and small dashing (green) line correspond to $w = (0.5,\, 1.0, \, 5.0)$ TeV cases, respectively. }
   \label{Fig:Couplings}
\end{figure}
%
%
%
%

Now, let us explore the fermion couplings in both the FC and FV scenarios, $\eta_{\mu}$ and $\eta_\tau$ and $\eta_{\mu \tau}$, respectively. In Fig.~\ref{Fig:MuTauMu}, we show
the magnitude of the couplings $\mu\mu$, $\tau\tau$, and $\mu\tau$ to the Higgs field. 
\begin{figure}[!htbp]
  \subfigure[]{ \includegraphics[width=7.7cm, height=5.3cm]{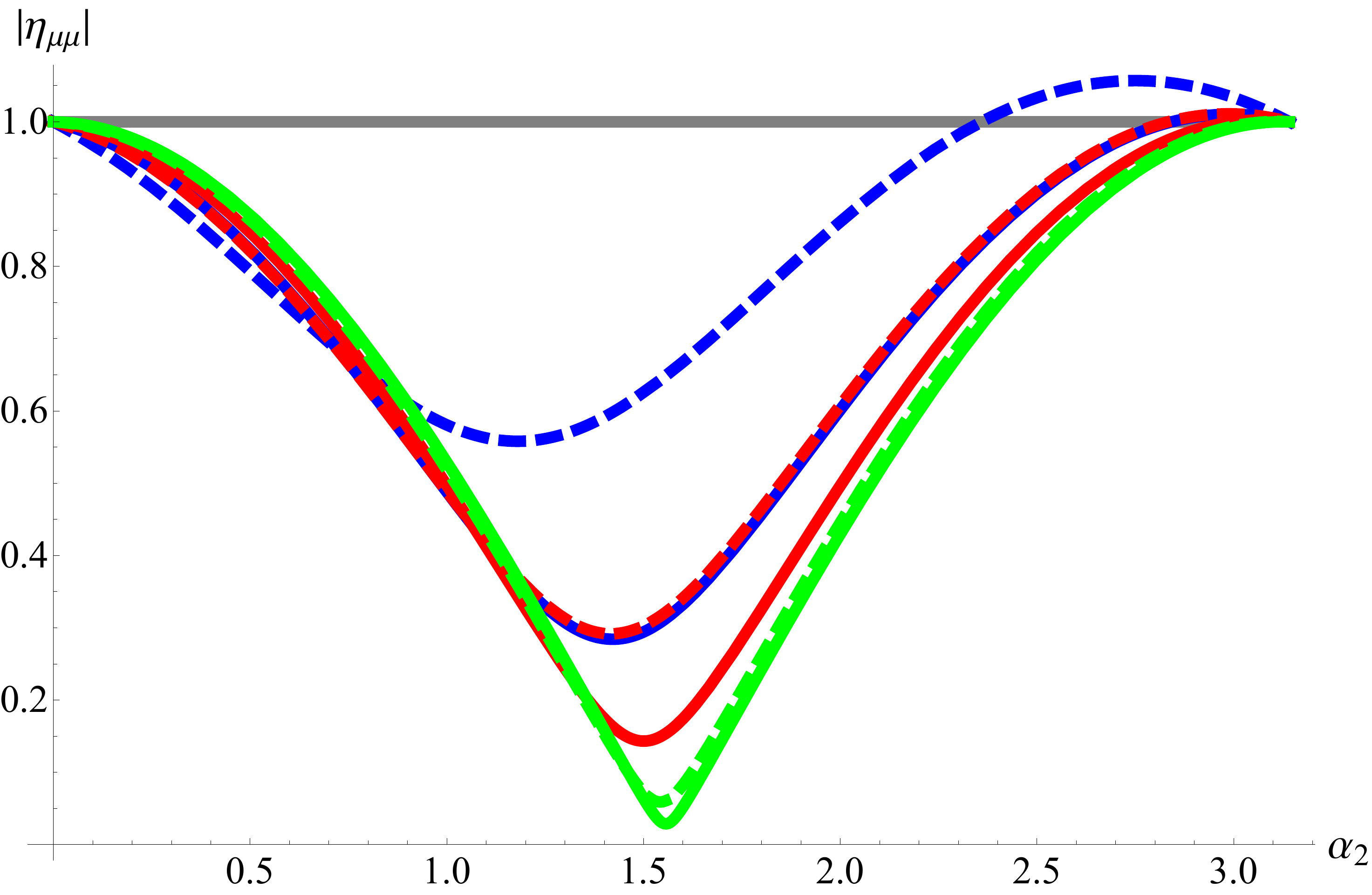}}
  \subfigure[]{ \includegraphics[width=7.7cm, height=5.3cm]{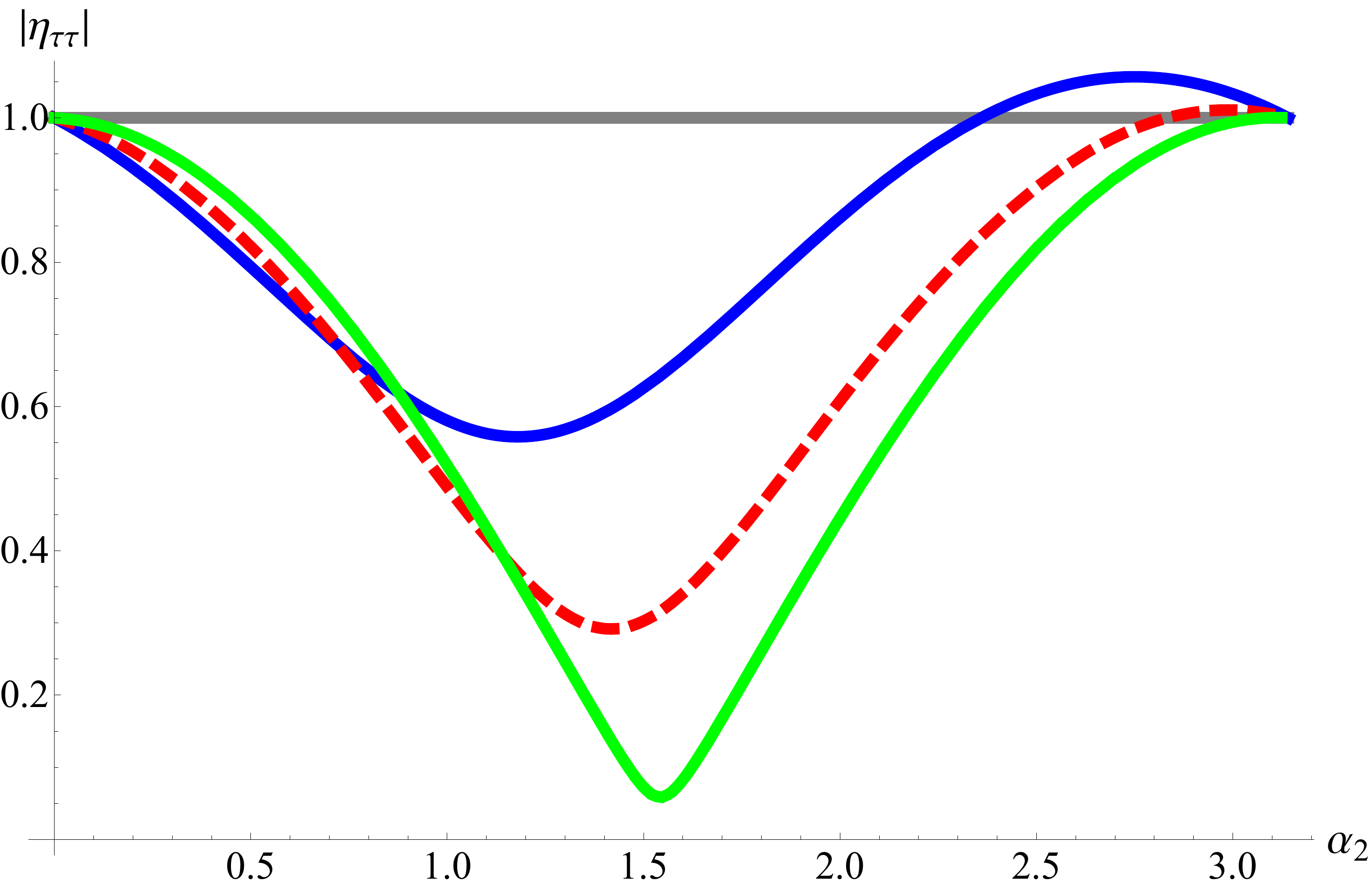}} \\
\subfigure[]{ \includegraphics[width=7.7cm, height=5.3cm]{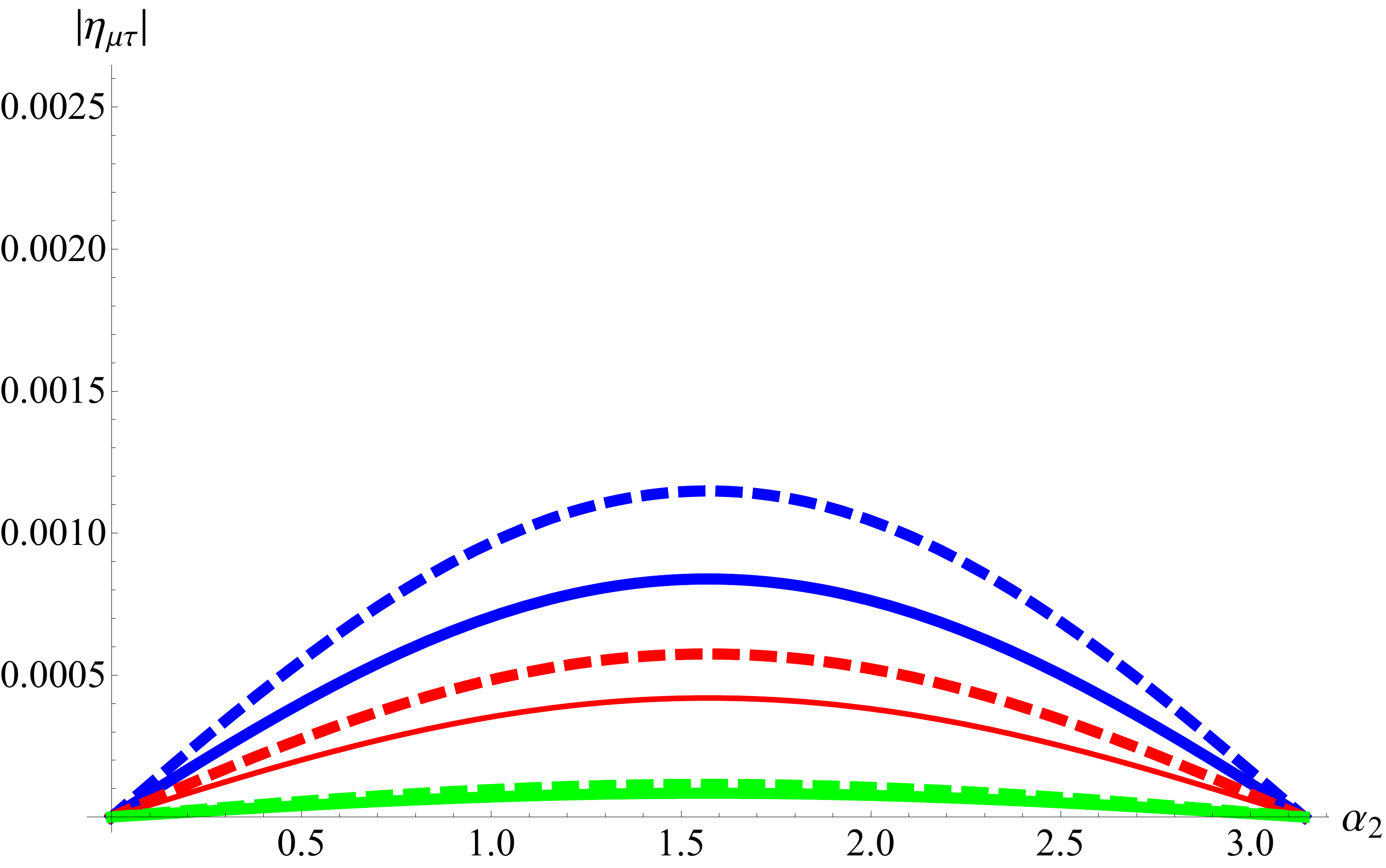}} 
  \caption{
The FC and FV Yukawa couplings, $\eta_{\mu \mu}$, $\eta_{\tau \tau}$, and $\eta_{\mu \tau}$, respectively, are shown against the $\alpha_2$ angle. It is being assumed that $\xi = \pi/7$ and $j=2$. The three different colors (blue, red, and green) correspond to $w = (0.5,\, 1.0, \, 5.0)$ TeV cases, respectively. On the other hand, the two different kinds of line: continuous and dashing, represent the two different values of $n$ used, $n=0$ and $n=2$, respectively, which basically tells us how the angle diagonalizing the mass matrix in the 2-3 sector relates to the mass ratio $m_\mu / m_\tau$, $\theta \simeq m_\mu / m_\tau$ and $\theta \simeq \sqrt{m_\mu / m_\tau}$, correspondingly. The (gray) horizontal line shows the SM value for the FC cases. }
   \label{Fig:MuTauMu}
\end{figure}
\subsection{LFV constraints from low-energy\label{subsec:LFVconst}}
When the small masses of neutrinos are included in the SM, charged lepton decays producing LFV are allowed starting the one loop level. Moreover, due to the smallness of neutrino masses compared to the mass of the charged weak bosons the probability of occurrence for these processes is, in fact, extremely small \cite{Deppisch:2012vj},
\begin{eqnarray}
	{B}r (\ell_i \rightarrow \ell_j \gamma) \sim \frac{\Delta m_\nu^4}{m_W^4}  \leq 10^{-54},
\end{eqnarray}
far below the present and foreseeable future resolution of experiments. In this sense, any finding of NP could be tracked via this kind of events as, commonly, extensions of the SM imply a branching ratio of $10^{-7}$. Therefore, they can provide us with a feasible way to look for deviations from the SM. Note that, already, the non-observation of these processes is giving strong bounds on BSM physics.

Among the charged lepton decays two of them, related to the muon and the tau, are of utmost importance. The former because NP could be responsible for the anomaly in the magnetic moment of the muon.  Whereas, the latter, because there are huge statistics collected by $BABAR$ \cite{Aubert:2009ag} and Belle \cite{Miyazaki:2011xe}, and also, recently, by the LHCb collaboration \cite{Aaij:2013fia}.

In the following, we want to study the consequences of our model in this set of decays.  In order to derive constraints on the LFV Higgs couplings, we shall use the tau decays  $\tau \to \mu \gamma$, $\tau \to 3\mu$, and the anomalous magnetic moment of
the muon.

The expressions for the  $\tau$ decay widths are~\cite{Harnik:2012pb},
\begin{eqnarray}
 \Gamma (\tau \to \mu \gamma) = \frac{\alpha m_\tau^5}{64\pi^4} \left( |c_L|^2 + |c_R|^2 \right), \\
 \Gamma (\tau \to  3\mu) \simeq \frac{\alpha^2 m_\tau^5}{6(2\pi)^5}\Bigg| \log \frac{m_\mu^2}{m_\tau^2}  - \frac{11}{4} \Bigg| \left( |c_L|^2 + |c_R|^2 \right),
\end{eqnarray}
where, 
\begin{eqnarray}
	c_L^{\text{1loop}} \simeq  \frac{1}{12m_h^2} \eta_{\tau \tau} \eta^*_{\tau \mu} \left(-4 + 3\log \frac{m_h^2}{m_\tau^2} \right), \quad
		c_R^{\text{1loop}} \simeq  \frac{1}{12m_h^2} \eta_{\tau \tau} \eta_{\mu \tau} \left(-4 + 3\log \frac{m_h^2}{m_\tau^2} \right).
\end{eqnarray}
These expressions are only valid within the hierarchical approximations $m_\mu \ll m_\tau \ll m_h$ and $\eta_{\mu \mu} \ll \eta_{\tau \tau}$. Moreover, the 2-loop contributions are also known and given numerically by,
\begin{eqnarray}
	c_L^{\text{2loop}} = \eta_{\tau \mu}^* (-0.082 \eta_{tt} + 0.11) \frac{1}{m_h^2}, \quad
	c_R^{\text{2loop}} = \eta_{\mu\tau} (-0.082 \eta_{tt} + 0.11) \frac{1}{m_h^2}.
\end{eqnarray}
By virtue of these equations, the following constraints have been obtained~\cite{Harnik:2012pb},
\begin{eqnarray}
	\sqrt{|\eta_{\mu \tau}|^2 + |\eta_{\tau \mu}|^2 } < 0.016, \\
	\sqrt{|\eta_{\mu \tau}|^2 + |\eta_{\tau \mu}|^2 } \lesssim 0.25, 
\end{eqnarray}
the former one using the tau decay $\tau \to \mu \gamma$ while the latter through the other decay mode $\tau \to 3\mu$. These constraints are computed by assuming that the FC Yukawa couplings are equal to the SM values~\cite{Harnik:2012pb}. In order to be valid for us these results, we need to restrain ourselves to the region $\alpha_2 < 0.4$ which is consistent with our previous finding. In 
Fig.~\ref{Taus}, we show how our model satisfies the previous constraints~\cite{Harnik:2012pb}. 

\begin{figure}[H]
\centering
\includegraphics[scale=0.35]{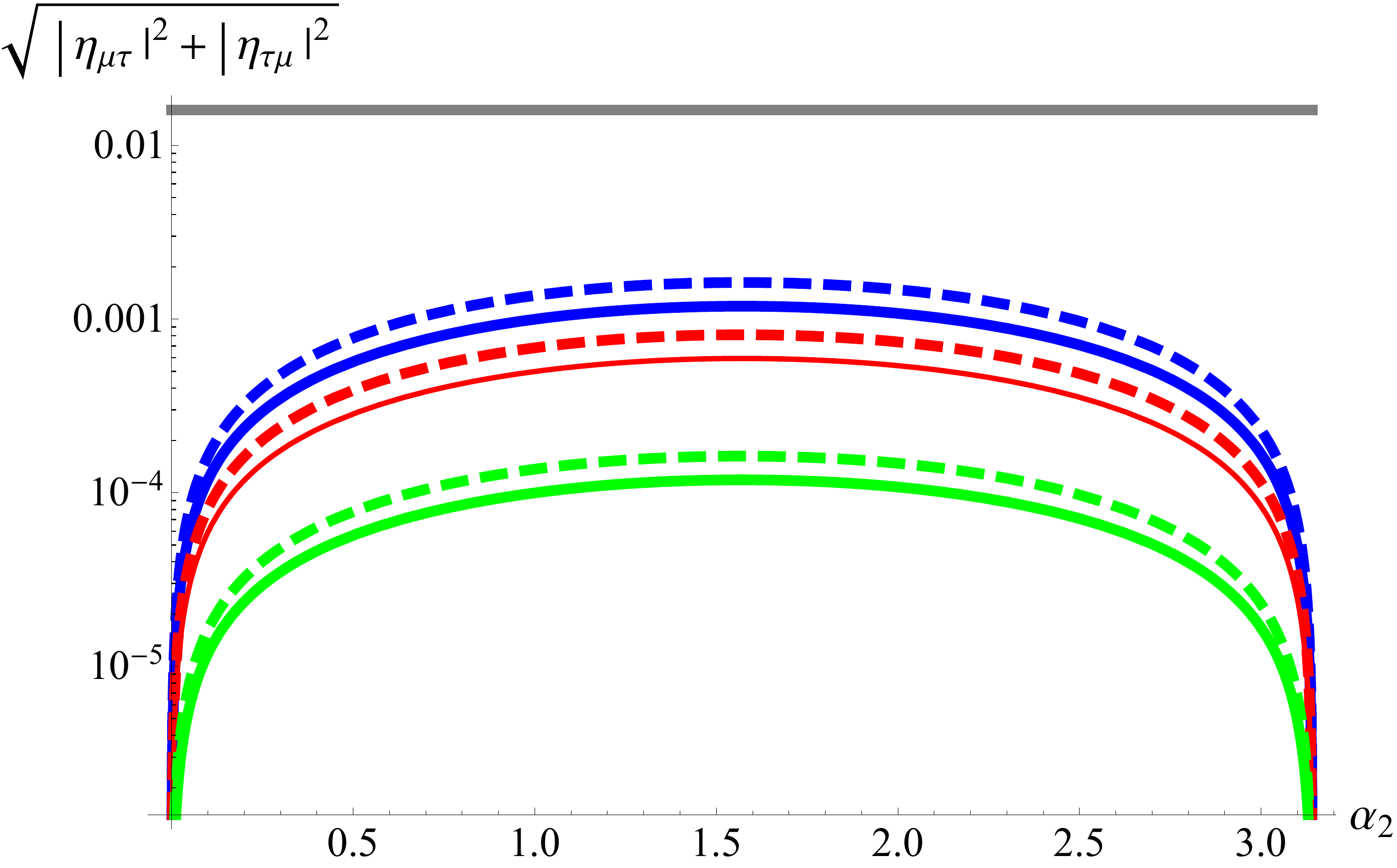}
\caption{The (gray) horizontal line shows the upper bound $\sqrt{|\eta_{\mu \tau}|^2 + |\eta_{\tau \mu}|^2 } < 0.016$ obtained in~\cite{Harnik:2012pb}. The three different colors (blue or upper, red or medium, and green or bottom) correspond to $w = (0.5,\, 1.0, \, 5.0)$ TeV cases, respectively. On the other hand, the two different kinds of line: continuous and dashing, represent the two different values of $n$ used, $n=0$ and $n=2$, respectively, which basically tells us how the angle diagonalizing the mass matrix in the 2-3 sector relates to the mass ratio $m_\mu / m_\tau$, $\theta \simeq m_\mu / m_\tau$ and $\theta \simeq \sqrt{m_\mu / m_\tau}$, correspondingly. Our model therefore satisfies all the experimental constraints.} 
\label{Taus}
\end{figure}

The present situation for the muon anomalous magnetic moment is still unclear as what is originating the observed discrepancy between the theoretical value from the experimental one,
\begin{eqnarray}
	\Delta a_\mu = a_\mu^{\text{exp}} - a_\mu^{\text{th}} = (261 \pm 80) \times 10^{-11},
\end{eqnarray}
which amounts to 3.3 standard deviations.  Nonetheless, the new Fermilab E989 experiment aims to improve the precision by a factor of four reducing the total uncertainty from 540 parts per billion to 140 parts per billion~\cite{Chapelain:2017syu}. From the theoretical viewpoint, the known uncertainty is well controlled and nobody doubts this could be producing the anomaly. It is even more likely that a mistake on the experimental side is being responsible for this or that several ingredients appearing in the theory predictions are not fully understood and possibly correlated~\cite{Melnikov:2016wdt, Melnikov:2006sr,Melnikov:2001uw}. About all, the more desirable and expected explanation to this anomaly is some physics beyond the SM~\cite{Czarnecki:2001pv}. In this respect, we now calculate the contribution coming from this model.
The expression for the anomalous magnetic moment we shall use is~\cite{Harnik:2012pb},
\begin{eqnarray}
     a_\mu \equiv \frac{g_\mu -2 }{2} \simeq \frac{\text{Re}(\eta_{\mu \tau} \eta_{\tau \mu}) }{8\pi^2} 
	\frac{m_\mu m_\tau}{2m_h^2} \left( 2 \log \frac{m_h^2}{m_\tau^2} - 3 \right).
\end{eqnarray} 
{In Fig.~\ref{muonAnomaly}, we show how the contribution coming from this model does not solve the anomaly. In fact, as already seen in \cite{Harnik:2012pb},  the requirement that should be met in order to solve the anomaly is given by,
\begin{eqnarray}
	{\text{Re}}(\eta_{\mu \tau} \eta_{\tau \mu}) \simeq (2.7 \pm 0.8) \times 10^{-3},
\end{eqnarray}
however, this is in conflict with the Cheng-Sher ansatz which implies, $| \eta_{\mu \tau} \eta_{\tau \mu} | \leq3.1 \times 10^{-6}$, so one needs to soften this restriction\footnote{In fact, by generalizing the Cheng-Sher ansatz is another approach by which this anomaly could be solved.}. }
\begin{figure}[H]
\centering
\includegraphics[width=7.7cm, height=5.3cm]{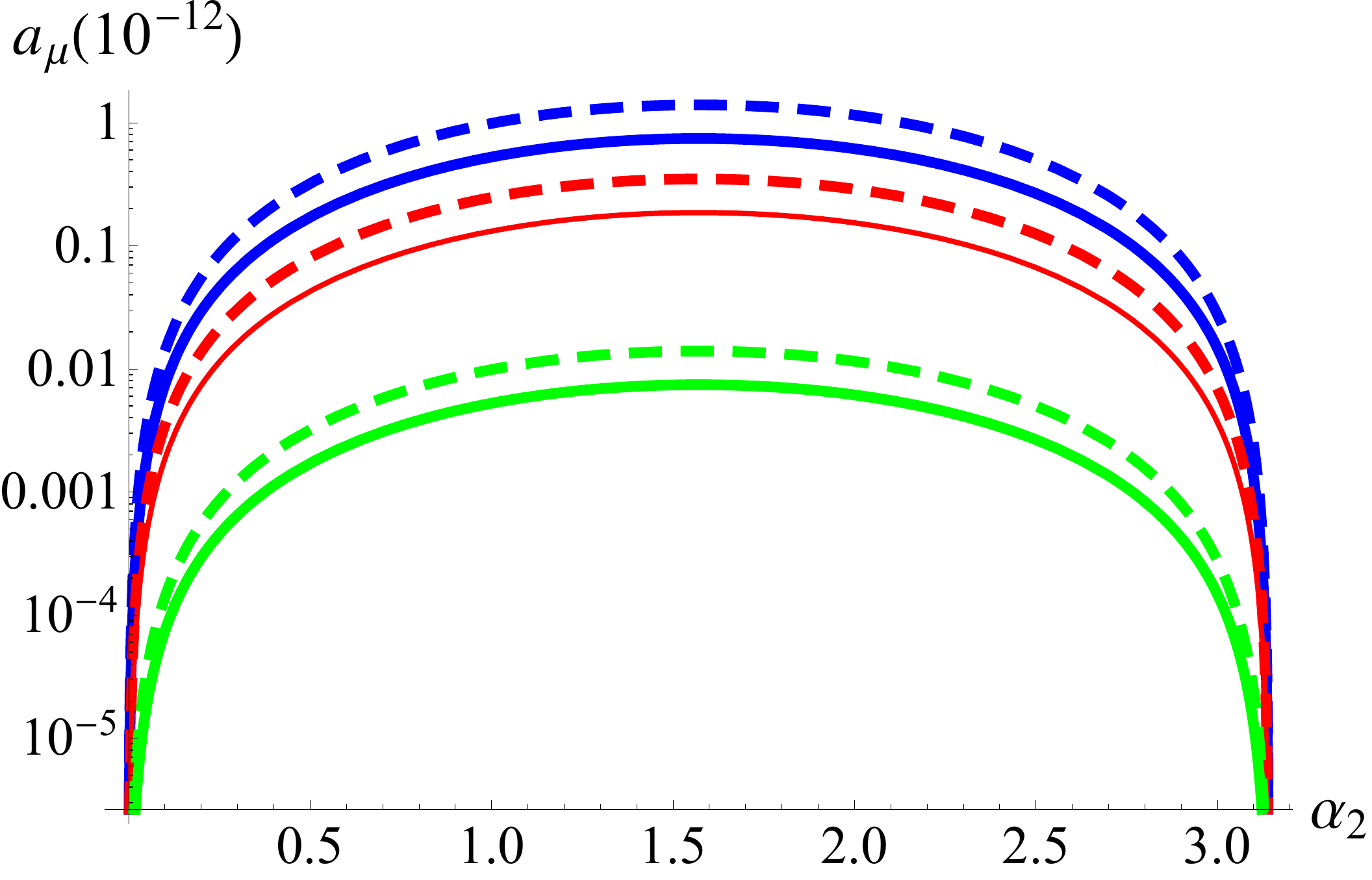}
\caption{Here it is shown the contribution to the anomalous magnetic moment of the muon. We see that we cannot solve the anomaly as a consequence of this model. 
The three different colors (blue or upper, red or medium, and green or bottom) correspond to $w = (0.5,\, 1.0, \, 5.0)$ TeV cases, respectively. On the other hand, the two different kinds of line: continuous and dashing, represent the two different values of $n$ used, $n=0$ and $n=2$, respectively, which basically tells us how the angle diagonalizing the mass matrix in the 2-3 sector relates to the mass ratio $m_\mu / m_\tau$, $\theta \simeq m_\mu / m_\tau$ and $\theta \simeq \sqrt{m_\mu / m_\tau}$, correspondingly.} 
\label{muonAnomaly}
\end{figure}


\subsection{LFV Higgs decay\label{Sec:LFV-H-decays}}
We shall present here predictions for the LFV Higgs decay as part of the test of the model using the relation given by $Br(h \to \tau\mu)$ as,
\begin{eqnarray}
Br(h \to \tau\mu) \simeq \frac{\Gamma(h\to \tau\mu)}{\Gamma (h \to \tau\tau)} Br_{SM}(h \to \tau\tau).
\end{eqnarray}
where $Br_{SM}(h \to \tau\tau)= 6.27\times10^{-2}$~\cite{Olive:2016xmw}. According to Eq.~(\ref{eq:casesvertex}), we must consider the two less suppressed cases, $n=0,2$, for the coupling $g_{h\mu\tau}$.

Numerical results are shown in Fig.~\ref{Fig:brmutaumodel}. For that, we consider the particular case (\ref{eq:particularcase})  and the parameter space 
$0.1<\lambda_s,\lambda_{11},\lambda_{12},\lambda_{s1},\lambda_{s2}<0.5$, $0.45<\tilde{\lambda}_{sa}<0.55$, 
$\lambda_{1}=0.5, \tilde{\lambda}_{sc}=0.1,\tilde{\lambda}_{sb}=1.5$, 
$\omega=0.5\,{\textrm{TeV}}$ and $124\,\textrm{GeV}<M_{h_1}<126\,\textrm{GeV}$.
\begin{figure}[!htbp]
\includegraphics[width=0.8\textwidth]{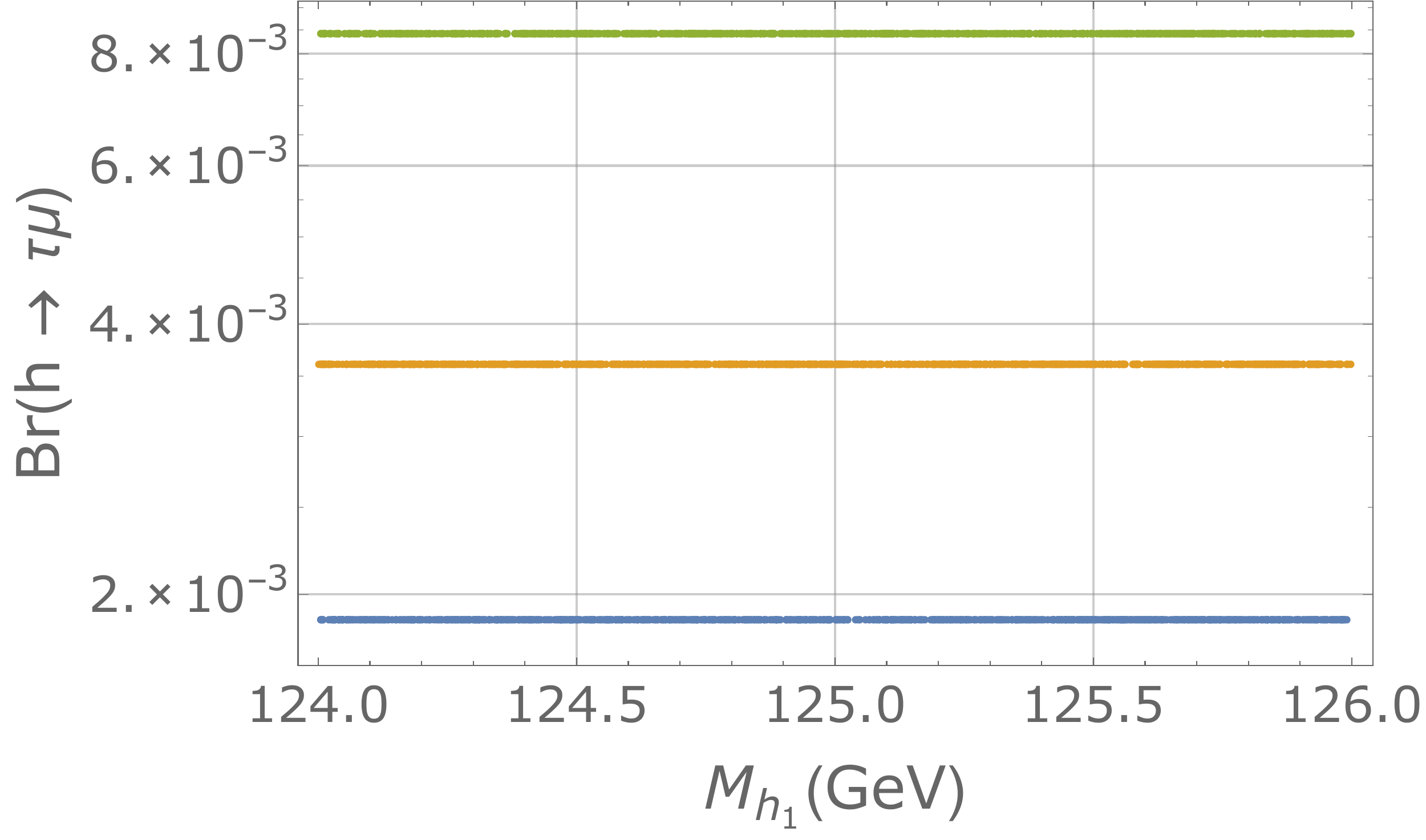}
  \caption{The experimental value of the branching ratio $Br(h \to \mu\tau)= 0.84^{+0.39}_{-0.37}\times10^{-2}$ (green line) is shown together with the theoretical values of 
 our model consisting of the SM plus a FN singlet for the two cases $n=0$ (blue line) and $n=2$ (orange line), respectively.}
   \label{Fig:brmutaumodel}
\end{figure}
Thus, in Fig.~\ref{Fig:brmutaumodel} we show the branching ratio $Br_{SM}(h \to \mu\tau)= 0.84^{+0.39}_{-0.37}\times10^{-2}$ (green line), and the associated results to our model with $n=0 \; (2)$ through the numerical values $Br(h \to \mu\tau)= 1.8736\times10^{-3} \;(3.6143\times10^{-3})$ in blue (orange) lines. These one show  the good agreement of the proposed model (SM with a FN singlet) with the experimental data related with NP reported in the literature and also its feasibility of being measured in the near future.

\section{Conclusions\label{Sec:Conclusions}}
In this paper, we have studied the LFV decays of the Higgs boson,  $h\to \ell_i \ell_j$, 
which vanish within the SM and are highly suppressed in several theoretical extensions.  This signal
is relatively simple to reconstruct at future colliders, and therefore, has become an important tool to probe 
SM extensions where these decays reaches detectable levels. We have identified a mechanism to induce LFV interactions
for the light SM-like Higgs boson,  by linking it with the appearance of CP violation in the scalar sector.
We have studied this idea first within the context of general multi-Higgs models, supplemented with a complex 
singlet.
This singlet scalar field is employed to provide an effective description of the fermion mass hierarchy, 
$m_3 \gg m_2 \gg m_1$, by incorporating the so called Froggatt--Nielsen (FN) mechanism. Moreover, 
by assigning it a complex vev, CP is spontaneously violated.
We have studied the consequences of such a model in producing lepton flavor violation 
(LFV) transitions via Yukawa interactions with the Higgs field, which are described by
an effective Lagragian that supports the generality of our mechanism. 

Then, in order to study Higgs phenomenology
we have focused in a minimal model, with an scalar sector  
consisting only of a Higgs doublet and a (complex) FN singlet. 
Then the scalar spectrum and the Yukawa interactions of the model were studied thoroughly. 
For that, we have first studied in all its details the 
scalar potential formed by all those new contributions coming from the two scalars of the theory. 
There, we have found through a scanning of the parameter space the following mass hierarchy for the neutral scalar states:
$M_{h1}^2 \ll M_{h3}^2 \ll M_{h2}^2$, where the lightest state is identified as the SM-like Higgs. 
We see that the ratio between the two heavy states decreases proportional to the amount of CP violation 
generated through the phase $\xi$ appearing in the complex flavon vev. When CP is conserved in the scalar 
potential the mass spectra gives almost two degenerate masses (125 GeV and 150 GeV) and one state much heavier 
than these two. 

Next, we considered the effective Yukawa Lagrangian. After taking both the fermion and the neutral 
component of the scalar fields to their corresponding mass eigenstates, we have
explicitly written the new couplings with the SM-like Higgs field (the lightest scalar state);
from which two kinds are identified: Flavor Conserving (FC) and Flavor Violating (FV). 
In order to generalize our study, we took the most generic features of leptonic models using a FN singlet. By virtue of them, we were able to provide, in a simple picture, the correct order of magnitudes for both the FC and FV Yukawa couplings within these models. The Cheng-Sher ansatz was also taken into account and showed to be a condition too strong to solve, in general terms, the anomalous magnetic moment of the muon.

Constraints on the parameters of the model are derived from low-energy
observables and LHC Higgs data, which is then applied to study the resulting predicted rates for the decay 
$h\rightarrow \tau \mu$. 
Overall, branching ratios for $h \rightarrow \tau \mu$ of the order $10^{-3}$ are obtained within this approach
consistent with all known constraints, which are well below the current bounds from the LHC,
i.e. $Br(h \to \mu  \tau ) \leq 10^{-2}$.

\begin{acknowledgments}
This work has been partially supported  by \textit{CONACYT-SNI (M\'exico)}. 
JLDC and UJSS want to acknowledge support by CONACYT-Mexico under Contract No. 220498. The authors thankfully acknowledge the computer resources, technical expertise and support provided by the ``Laboratorio Nacional de Superc\'omputo del Sureste de M\'exico through the grant number O-2016/039".
\end{acknowledgments}


\end{document}